\documentclass[10pt]{iopart}
\usepackage[T1]{fontenc}
\usepackage{txfonts}
\usepackage{geometry}
\usepackage{flushend}
\usepackage{fancyhdr}
\usepackage{iopams}
\usepackage{setstack}
\usepackage[ruled,vlined]{algorithm2e}
\usepackage{array}
\usepackage[font=small,labelfont=bf]{caption}
\usepackage[subrefformat=simple,labelformat=parens]{subfig}
\usepackage{url}
\usepackage{xcolor}
    \definecolor{dark-red}{rgb}{0.4,0.1,0.1}
    \definecolor{dark-blue}{rgb}{0.15,0.15,0.5}
    \definecolor{medium-blue}{rgb}{0.0,0.0,0.8}
    \definecolor{mycolor1}{rgb}{1,0.2,0.2}
\usepackage{booktabs}
\usepackage{cite}
\usepackage{dashrule}
\usepackage{fixltx2e}
\usepackage{graphicx}
\usepackage[activate={true,nocompatibility},final,tracking=true,kerning=true,spacing=true,factor=1100,stretch=10,shrink=10]{microtype}
\usepackage{float}
    \restylefloat{table}
\usepackage{tikz}
\usepackage{pgfplots}
\usepackage{hyperref}
    \hypersetup{
    colorlinks, linkcolor={dark-red},citecolor={medium-blue}, urlcolor={medium-blue},
    backref=true
    }

\usepackage{subeqn}
\graphicspath{{figs/}}

\definecolor{hwmycolor1}{rgb}{0.4941,0.5725,1}
\definecolor{hwmycolor2}{rgb}{1,0.7843,0.3922}
	
\makeatletter
\let\refP\@refstar
\makeatother
\newcommand{\plotref}[1]{\refP{#1}}

\newcommand{\wf}{\omega_{{\mathrm{f}}}}
\newcommand{\wfh}{\hat{\omega}_{{\mathrm{f}}}}
\newcommand{\kf}{\kappa_{{\mathrm{f}}}}
\newcommand{\lambdaf}{\lambda_{{\mathrm{f}}}}
\newcommand{\lambdaa}{\lambda_{{\mathrm{a}}}}
\newcommand{\alphaf}{\alpha_{{\mathrm{f}}}}

\newcommand{\supsc}[2]{{#1}$^{\mathrm{#2}}$}

\newcommand{\akh}{\hat{a}_k}
\newcommand{\phikh}{\hat{\phi}_k}
\usepackage{xspace} 
\newcommand{\SNRin}{SNR\textsubscript{in}\xspace}
\newcommand{\SNRout}{SNR\textsubscript{out}\xspace}
\newcommand{\fs}{f_\mathrm{s}}

\usepackage{dcolumn}
\usepackage{multirow}
    \newcolumntype{L}[1]{D{.}{.}{#1}}
    \newcolumntype{d}[1]{D{-}{\mbox{\,--\,}}{#1}}

\makeatletter
 \renewenvironment{table}
     {
      \renewcommand* {\@floatboxreset}{%
      \reset@font\@setminipage}
      \small\@float{table}
      }
     {\end@float\normalsize}
\makeatother

\fancypagestyle{firststyle}
{
   \fancyhf{}
   \fancyfoot[C]{\thepage}
}

\begin{document}
\twocolumn[ 
\title{\Huge A Fast, Robust Algorithm for Power Line Interference Cancellation in Neural Recording}

\noindent
\author{\large Mohammad Reza Keshtkaran and Zhi Yang}
\vspace{0.3cm}
\address{Department of Electrical and Computer Engineering
, National University of Singapore,\\
117583 Singapore
  }
\ead{\href{mailto:keshtkaran@nus.edu.sg}{keshtkaran@nus.edu.sg}}
    \maketitle

\begin{minipage}{0.78\textwidth}

\renewcommand{\abstractname}{\textbf{\normalsize Abstract}}
\begin{abstract}
\\ \noindent\normalsize
\emph{Objective}
Power line interference may severely corrupt neural recordings at 50/60~Hz and harmonic frequencies. The interference is usually non-stationary and can vary in frequency, amplitude and phase. To retrieve the gamma-band oscillations at the contaminated frequencies, it is desired to remove the interference without compromising the actual neural signals at the interference frequency bands.
In this paper, we present a robust and computationally efficient algorithm for removing power line interference from neural recordings.
\emph{Approach}
The algorithm includes four steps. First, an adaptive notch filter is used to estimate the fundamental frequency of the interference. Subsequently, based on the estimated frequency, harmonics are generated by using discrete-time oscillators, and then the amplitude and phase of each harmonic are estimated through using a modified recursive least squares algorithm. Finally, the estimated interference is subtracted from the recorded data.
\emph{Main results}
The algorithm does not require any reference signal, and can track the frequency, phase, and amplitude of each harmonic. When benchmarked with other popular approaches, our algorithm performs better in terms of noise immunity, convergence speed, and output signal-to-noise ratio (SNR).
While minimally affecting the signal bands of interest, the algorithm consistently yields fast convergence ($<$\,100~ms) and substantial interference rejection (output SNR $>$\,30~dB) in different conditions of interference strengths (input SNR from $-$30~dB to 30~dB), power line frequencies (45--65~Hz), and phase and amplitude drifts. In addition, the algorithm features a straightforward parameter adjustment since the parameters are independent of the input SNR, input signal power, and the sampling rate.
A prototype was fabricated in a \mbox{65-nm} CMOS process and tested. The MATLAB implementation of the algorithm has been made available for open access at \mbox{\url{https://github.com/mrezak/removePLI}}.
\emph{Significance} The proposed algorithm features a highly robust operation, fast adaptation to interference variations, significant SNR improvement, low computational complexity and memory requirement, and straightforward parameter adjustment. These features render the algorithm suitable for wearable and implantable sensor applications, where reliable and real-time cancellation of the interference is desired.
\end{abstract}
\end{minipage}
\hrule
\vspace{0.3cm}
] 
\section{Introduction}
Extracellular neural recordings have made it possible to monitor single-neuron and population activities for studying various cognitive and motor functions.
Due to various recording imperfections and experimental protocols, neural recordings are frequently superimposed with interferences and artefacts, which can cause erroneous data analysis. A more common cause of concern is the power line interference which is mainly due to the capacitive coupling between the subject and nearby electrical appliances and mains wiring \cite{chimene_comprehensive_2000,mitra_observed_2007}.

While high signal-to-noise ratio (SNR) (i.e.~ power of the clean neural signal divided by the power of the interference) is preferred for reliable data analysis, the interference pickup can be severe, degrading the SNR to as low as $-20$~dB (the interference is 100 times stronger than the signal).
This is especially the case in some experiments where the operation of nearby electrical appliances is unavoidable, and the desired recording isolations cannot be obtained \cite{rijn_high-quality_1990,chimene_comprehensive_2000,teplan_fundamentals_2002,mitra_observed_2007,thorp_interference_2009}.
\thispagestyle{firststyle} 

For studying field potentials at lower frequencies (e.g.~$<$\,30~Hz), a low-pass filter is sufficient to reject the power line interference. However, there is an increasing attention to the gamma band oscillations ($>$\,30~Hz) due to their correlation with a wide range of cognitive and sensory processes~\cite{buzsaki_high-frequency_1992,jones_intracellular_2000,staba_quantitative_2002,mitra_observed_2007,chao_long-term_2010,leuthardt_braincomputer_2004,miller_beyond_2008,moran_evolution_2010,shimoda_decoding_2012,liang_decoding_2012,hwang_utility_2013}. For example, the frequency bands of  \mbox{80--500~Hz} in \cite{staba_quantitative_2002}, \mbox{40--180~Hz} in \cite{leuthardt_braincomputer_2004}, \mbox{76--150~Hz} in \cite{miller_beyond_2008}, \mbox{0--200~Hz} in \cite{miller_decoupling_2009}, and \mbox{30--200~Hz} in \cite{liang_decoding_2012} have been shown useful for studying cognitive and motor processing. In this case, in addition to the fundamental harmonic at 50~Hz or 60~Hz, high order harmonics of the interference should also be removed before data analysis.

The interference is usually non-stationary and can vary in frequency, amplitude and phase. The frequency variations are usually small, and mainly originated from the AC power system \cite{dugan_electrical_2003,baggini_handbook_2008}. Nevertheless, the amplitude and phase variations can be large, which may significantly decrease the SNR of the recorded signal. These variations are mostly due to the subject movements, abrupt changes in nearby AC loads, and changes in capacitive coupling \cite{dugan_electrical_2003,chimene_comprehensive_2000,wang_trial--trial_2011}.
As a result, automatic cancellation of non-stationary power line interference would be advantageous for reliable data analysis.

	\begin{figure*}[t]
		\centering
		\includegraphics[width = \textwidth]{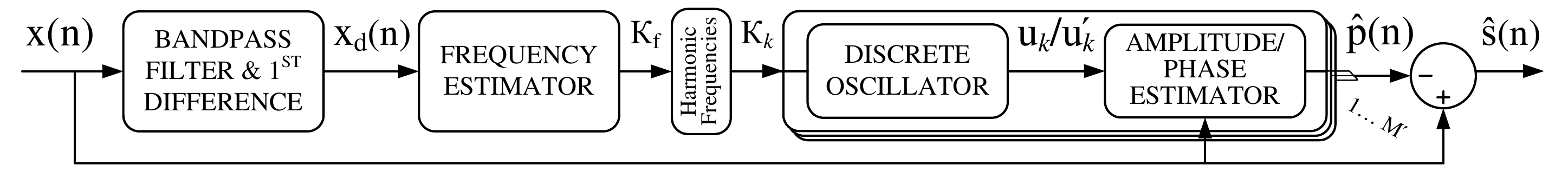}
		\caption{Functional block diagram of the proposed algorithm.  $x(n)$ is the input signal contaminated by power line interference, $\hat{p}(n)$ is the estimated interference, and $\hat{s}(n)$ is the output interference-free signal.
		}
		\label{fig:blockdiag}
	\end{figure*}
	
A number of solutions are available for reducing the interference pickup. To attenuate the interference at hardware level, biopotential amplifiers are frequently designed to take differential input with large common mode rejection ratio and large isolated-mode rejection ratio. In addition, using active electrodes, shielding electrodes and the subject, and grounding the nearby electrical appliances are useful ways to further reduce the interference ~\cite{rijn_high-quality_1990,chimene_comprehensive_2000,degen_enhancing_2004,spinelli_two-electrode_2005,alzaher_highly_2013}. Despite these considerations, large residual interference may remain in the signal~\cite{chimene_comprehensive_2000,ziarani_nonlinear_2002,mitra_observed_2007,thorp_interference_2009}, thus further signal processing is required to completely remove the interference.

Notch filtering has been widely used to attenuate the interference by rejecting its predetermined frequency components (i.e.~at 50/60~Hz and harmonic frequencies). To avoid making excessive distortion, the filter should feature narrow notch bandwidth, small phase distortion, and negligible artificial oscillations~\cite{wang_trial--trial_2011, martens_improved_2006, luck_introduction_2005,levkov_removal_2005}.
However, it is difficult to meet these specifications when the interference frequency is not stable and the filter is to accommodate the frequency variations. On the one hand, a very narrow notch may lead to an inadequate removal of the interference, especially when its frequency shifts outside the notch bandwidth. On the other hand, a wide notch can attenuated the interference, but it also results in the excessive removal of information-bearing signal components. These reasons have made notch filtering not a good candidate for power line interference removal in neural recording applications \cite{mitra_observed_2007,wang_trial--trial_2011}.

Other techniques based on spectrum estimation have been used for detecting and removing the spectral peaks (thus the interference) \cite{mitra_observed_2007}. A drawback is that, these methods require buffering a large number of samples, which slows down the signal processing and is not suitable for \mbox{real-time} implementation. Furthermore, they usually lose their effectiveness when the interference is non-stationary \cite{ziarani_nonlinear_2002,levkov_removal_2005}.

Another popular approach is to use adaptive interference cancellation which addresses some of the drawbacks of notch filtering. When an auxiliary reference signal of the interference is available, the well-known adaptive noise canceller (ANC) can be utilized to remove the interference \cite{widrow_adaptive_1975,hamilton_comparison_1996,wang_trial--trial_2011}. However, it may become ineffective when the interference contains higher order harmonics. Moreover, a reference signal may not always be available in practice.
To address these limitations, several reference-free adaptive methods have been proposed, mainly tailored for electrocardiography (ECG) signal processing  \cite{ferdjallah_adaptive_1994,ziarani_nonlinear_2002,levkov_removal_2005,martens_improved_2006}. Nevertheless, the performance and reliability of these methods have not been tested on neural recordings. In general, several issues might arise when applying the same algorithms to neural recordings. For example, in some algorithms \cite{martens_improved_2006,levkov_removal_2005}, the detection of QRS periods of the ECG signals is necessary to tackle non-stationarity; however, this method is not applicable to neural signals since the on/off period of neural oscillations cannot be easily detected in the presence of the interference. In addition, the power spectral density (PSD) of neural signals follows \mbox{$1/f^\alpha${\scriptsize$(1\,{<}\,{\alpha}\,{<}\,3)$}} distribution \mbox{\cite{heldman_local_2006,miller_spectral_2007,miller_decoupling_2009}} which is different from the that of the ECG; this might lead to inaccurate operation of the interference removal algorithms that are specifically tailored for ECG processing.

This paper proposes an algorithm which can reliably estimate and remove the 50/60~Hz line interference and its harmonics from neural recordings. The algorithm does not require any reference signal, and can track the variations in the frequency, phase, and amplitude of the interference at both the fundamental and the harmonic frequencies. The algorithm can reject the interference, while minimally affecting the signal band of interest, achieving output SNR (i.e.~SNR after interference cancellation) of over 30~dB. When applied to neural signals, a performance comparison with two adaptive methods of {\cite{ziarani_nonlinear_2002}} and {\cite{martens_improved_2006}} is carried out, where the proposed algorithm outperforms in terms of convergence behaviour and output SNR. The low computational complexity, low memory requirement, and adequate numerical behaviour of the algorithm make it suitable for \mbox{real-time}, low-latency hardware implementation. The algorithm is implemented in software as well as an application-specific integrated circuit (ASIC). The software source code and its user manuals are available for open access at {\cite{MATLABCode}}. The ASIC was fabricated in a 65-nm CMOS process, and its robust and real-time operation is verified.
A preliminary version of this work has been presented in \cite{keshtkaran_power_2012}.

The rest of this paper is organized as follows. \Sref{sec:methods} details the proposed algorithm, its pseudocode, and parameter adjustment. \Sref{sec:simulations} gives the experimental results based on both synthesized and real data, and presents a performance comparison with other methods. \Sref{sec:discussion} presents the discussion, and \sref{sec:conclusion} concludes the paper. The mathematical derivation of the algorithm are given in \ref{sec:apxMath}. The ASIC implementation and testing results are briefly described in \ref{sec:apxHardware}.

\section{Proposed Algorithm} \label{sec:methods}
A recorded neural signal from one electrode can be represented by
	\begin{eqnarray} \label{eq:sigmodel}
		x(n)=s(n)+p(n) , \quad n \in \mathbb{Z},
	\end{eqnarray}
where $x(n)$ is the measured signal, $s(n)$ is the signal of interest (neural signal + neural noise), and $p(n)$ is the power line interference, all sampled at $\fs$~Hz. $x(n)$ is assumed to be zero-mean, $s(n)$ has a $1/f^\alpha${\scriptsize$(1\,{<}\,{\alpha}\,{<}\,3)$} power spectrum, and $p(n)$ consists of a set of harmonic sinusoidal components with unknown frequencies, phases and amplitudes as
\begin{eqnarray}
    \eqalign{
	p(n) &= \sum\limits_{k=1}^M \underbrace{a_k \cos(k \wf n + \phi_k)}_{h_k(n)}
         = \sum\limits_{k=1}^M h_k(n). \label{eq:pli}
    }
\end{eqnarray}
Here, $\wf$ is the fundamental frequency in rad/s, $a_k$ and $\phi_k$ are the amplitude and phase of the $k^\mathrm{th}$ harmonic, and $M$ is the number of harmonics present in the interference.

An ideal interference cancellation algorithm should eliminate the  interference $p(n)$, while perfectly preserving the neural signal $s(n)$.
Let $\wfh$, $\akh$, $\phikh$, $\hat{h}_k(n)$, and $\hat{p}(n)$ denote the estimate of $\wf$, $a_k$, $\phi_k$, $h_k(n)$, and $p(n)$, respectively. The clean (i.e.~interference-free) signal $\hat{s}(n)$ is obtained as
\begin{subequations}
\begin{equation} 	
		 \hat{s}(n) = x(n) - \hat{p}(n),
\end{equation} 	
where
\begin{equation} 	
		 \hat{p}(n) = \sum\limits_{k=1}^{M'}\hat{h}_k(n).
\end{equation} 	
\end{subequations}
Here, $M'$ represents the desired number of harmonics to be removed from the recorded signal. It is chosen based on the bandwidth of interest, and its maximum value $M'_{\max} = \lfloor \pi / \wfh \rfloor$ can be adopted if it is desired to remove all the harmonics up to the Nyquist frequency.

The following approach is proposed to cancel the interference. First, the interference fundamental frequency $\wf$ is estimated by using a fast and numerically well-behaved frequency estimator.
Subsequently, based on the estimated frequency $\wfh$, each harmonic signal $h_k(n)$ is obtained by using discrete-time oscillators and then its amplitude and phase (i.e.~$\hat{a}_k$  and $\hat{\phi}_k$, respectively) are estimated by using a simplified recursive least squares (RLS) algorithm. The cascaded stages of frequency and amplitude/phase estimation allow individually adjustable adaptation rates for each of these estimators, which helps to achieve a fast and reliable estimation of the interference. Finally, the estimated interference $\hat{p}(n)$ is subtracted from the input signal $x(n)$ to obtain the clean signal $\hat{s}(n)$. The structure of the proposed algorithm is shown in \fref{fig:blockdiag}.

\subsection{Fundamental Frequency Estimation}
For robust estimation of the fundamental frequency, first, the signal is preprocessed to enhance the fundamental harmonic of the interference. After that, the enhanced signal is used for frequency estimation.
The preprocessing stage is described in \sref{sec:preprocess}, followed by the frequency estimation stage in \sref{sec:freqest}.

\subsubsection{Preprocessing: Initial Band-pass Filtering and Spectrum Shaping} \label{sec:preprocess}
\begin{figure}[t]
	\centering
	\captionsetup[subfigure]{oneside,margin={8mm,0mm}}
    \subfloat[][]{\includegraphics{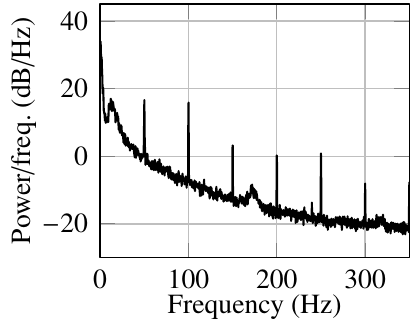}\label{fig:PSDshaping1}}
	\hfill
    \subfloat[][]{\includegraphics{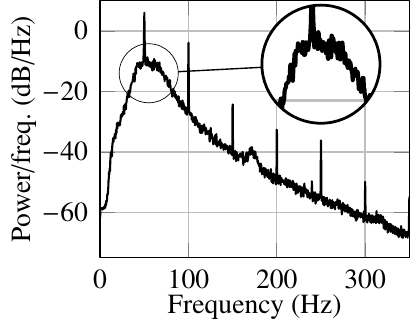}\label{fig:PSDshaping3}}
\caption{\label{fig:PSDshaping}The effect of bandpass filtering and spectrum shaping. (a) PSD of a real ECoG signal. (b) PSD after bandpass filtering and spectrum shaping, where the fundamental harmonic is enhanced.}
\end{figure}

Since the input signal $x(n)$ has a coloured PSD ($1/f$), the direct application of a typical adaptive frequency estimator would lead to a biased estimation of the frequency \cite{nam_ik_cho_adaptive_1989}. It is also possible that the inference $p(n)$ is weak at its fundamental frequency and more dominant at certain harmonic frequencies, especially with the use of a differential recorder that can largely attenuate \mbox{odd-order} harmonics. This may prevent the frequency estimator from converging to a correct frequency estimate.
To address these issues and improve the frequency estimation, the input signal $x(n)$ is bandpass filtered with a \mbox{4\textsuperscript{th}-order} infinite-impulse-response (IIR) filter to enhance the fundamental harmonic of the interference and attenuate higher harmonics. This filtering is also useful for attenuating lower frequency artefacts and signal components which may negatively affect the frequency estimation.
The filter passband is by default set to 40--70~Hz to accommodate both 50~Hz and 60~Hz power line frequencies and their worst case variations, but it can be further customized; for example, to 55--65~Hz if the nominal power line frequency is known to be 60~Hz.
Let $H(\cdot)$ be the realization of the bandpass filter, the filtered  signal $x_f(n)$ is obtained as
\begin{subequations}
\begin{eqnarray}
  x_\mathrm{f}(n) = H(x(n)).
\end{eqnarray}

To further reduce the estimation bias, a \mbox{1\textsuperscript{st}-order} differentiator is utilized which mitigates the effect of the power law spectrum of the input signal:
\begin{eqnarray}
 x_\mathrm{d}(n) = x_\mathrm{f}(n) - x_\mathrm{f}(n-1).
\end{eqnarray}
\end{subequations}
Here, $x_d(n)$ is the first difference signal fed into the next stage ANF for frequency estimation. The effect of \mbox{1\textsuperscript{st}-order} differentiation on the overall performance of the algorithm is not very significant; however, in practice, the first order differentiator can be incorporated into the bandpass filter with negligible computational overhead. \Fref{fig:PSDshaping} shows the effects of bandpass filtering and spectrum shaping, where the fundamental harmonic of the interference is enhanced.
It should be noted that signal $x_\mathrm{d}$ is only used for frequency estimation, and not for amplitude/phase estimation.

\begin{figure*}[t]
	\subfloat[][]{\label{fig:lattice}\includegraphics[width =5cm]{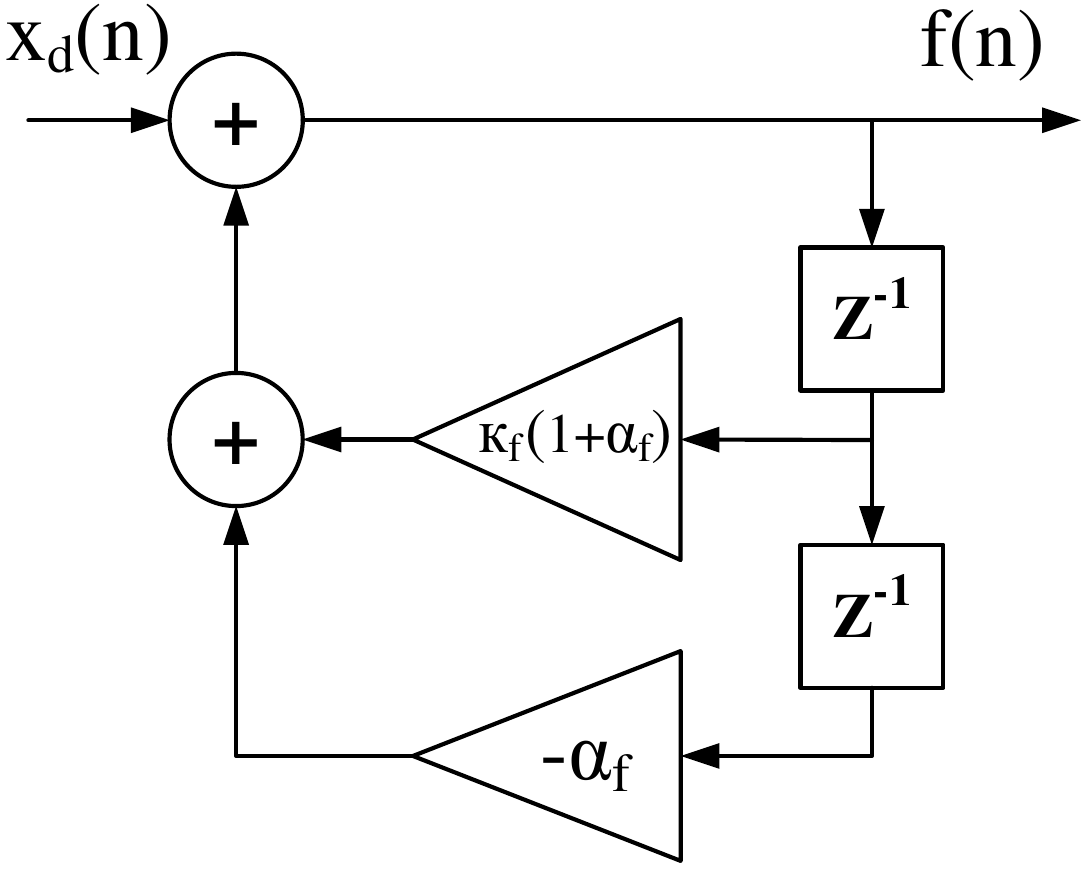}}
	\hfill
	\subfloat[][]{\label{fig:oscillator}\includegraphics[width =5cm]{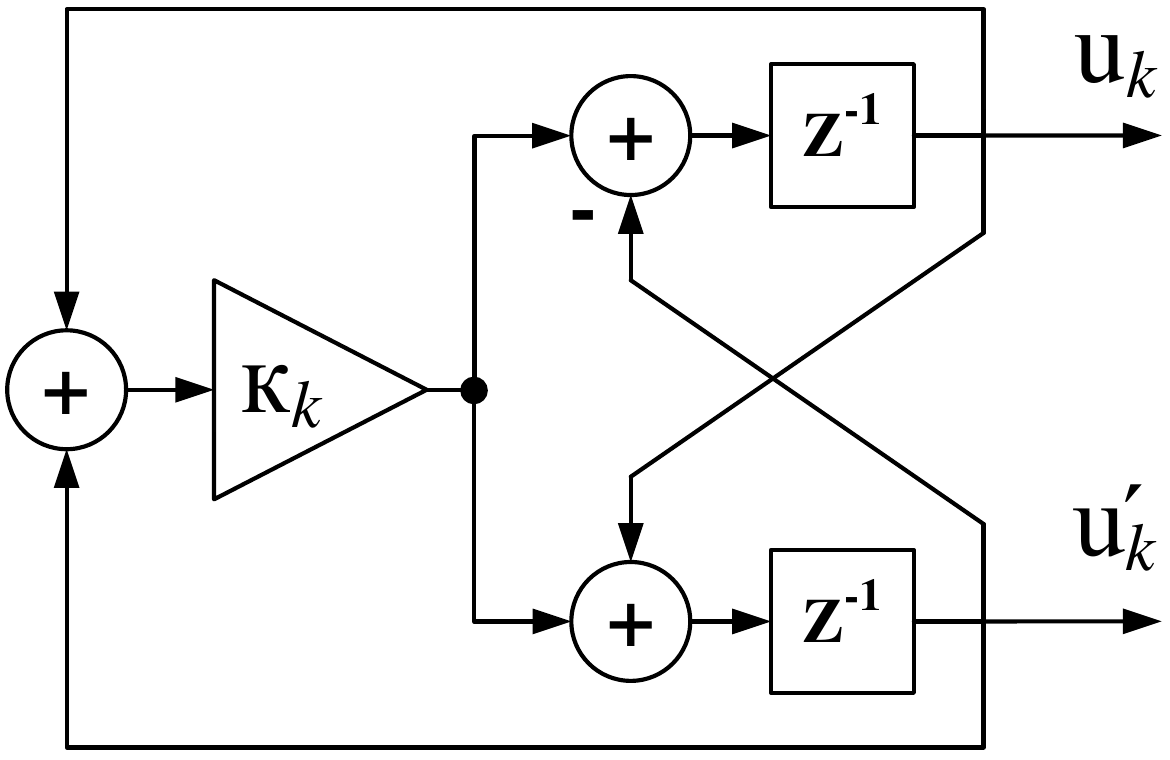}}
	\hfill
	\subfloat[][]{ \label{fig:combiner}	\includegraphics[width =4cm]{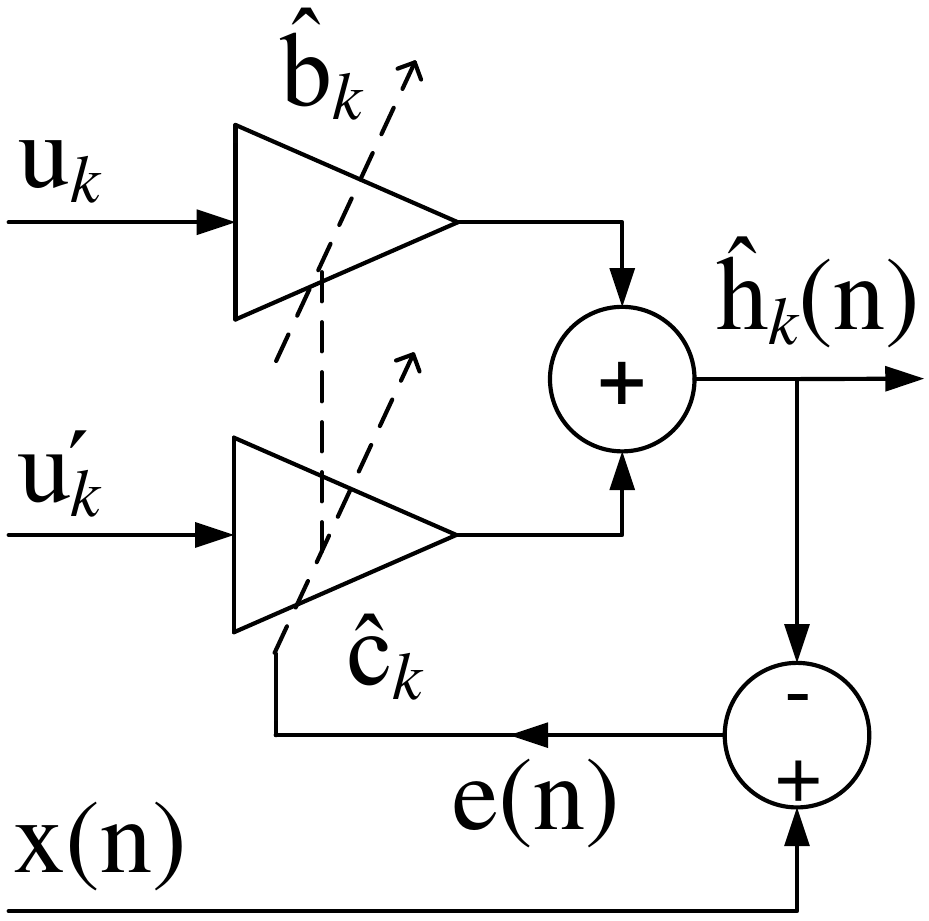}}
	\caption{Signal flow graph of (a) all-pole lattice ANF structure. Notch frequency and bandwidth are determined by $\kf$ and $\alphaf$, respectively. (b) Discrete-time oscillator. The parameter $\kappa_k$ adjusts the oscillation frequency. $u_k$ and $u'_k$ represent orthogonal sinusoids at frequency $k\wf$. (c) Adaptive linear combiner, used for amplitude/phase adaptation. The weights $a_k$ and $b_k$ are adapted by the simplified RLS algorithm which minimizes the weighted least square error between $x(n)$ and $\hat{h}_k(n)$.}
\end{figure*}

\subsubsection{Frequency Estimation} \label{sec:freqest}
The estimation of the instantaneous frequency of a single sinusoid buried in broadband noise has been largely investigated in the literature. Various well-established methods exist for frequency estimation differing in performance with regard to computational complexity, and estimation bias and variance \cite{nehorai_minimal_1985,kay_fast_1989,regalia_stable_1992,cho_adaptive_1993,klein_fast_2006}. In this work, a lattice adaptive notch filter (ANF)-based frequency estimator is utilized since it features instantaneous estimation of the frequency, desirable performance, low complexity, and suitability for real-time finite-precision implementation.

It should be noted that the ANF is merely used for frequency estimation and not for notch filtering, hence the \mbox{all-zero} section need not be used.
\Fref{fig:lattice} shows the structure of the ANF where $x_d(n)$ is the input signal from the preprocessing stage and $f(n)$ is the output of the \mbox{all-pole} section. The transfer function of the \mbox{all-pole} section is given by
\begin{eqnarray}
	H(z) =  \frac{1}{1 - \kf(n)(\alphaf+1)z^{-1} + \alphaf z^{-2}}, \label{eq:trans}
\end{eqnarray}
where  $\kf(n)$ is the adaptive coefficient at time step $n$, which gives the frequency estimate $\wfh(n)$ through $\wfh(n)=\cos^{-1}\kf(n)$, and 0\;$<$\;$\alphaf$\;$<$\;1 is the pole radii and determines the notch bandwidth.
The lattice algorithm of \cite{cho_adaptive_1993} is employed to adjust $\kf$ as follows.
\begin{subequations}
 \begin{eqnarray}
	\label{eq:latticeAlgo}
    c(0) &= d(0) = \epsilon >0, \; f(-1)=f(-2)=0,\; \kf(0)=0, \label{eq:latticeinit}\\
	c(n) &= \lambdaf c(n-1)+f(n-1)(f(n)+f(n-2)),\label{eq:latticec} \\
	d(n) &= \lambdaf d(n-1) + 2 f(n-1)^2, \\
	{\kappa}_\mathrm{t}(n) &= \frac{c(n)}{d(n)}, \label{eq:latticek}	\\
    {\kappa}_\mathrm{t}(n) &= \left\{
    \begin{array}{rcl}
        {\kappa}_\mathrm{t}(n) & \quad \hbox{if} \quad & -1 < {\kappa}_\mathrm{t}(n) < 1, \\
        1 & \quad \hbox{if} \quad & {\kappa}_\mathrm{t}(n) > 1, \\
        -1 &\quad \hbox{if} \quad& {\kappa}_\mathrm{t}(n) < -1,
    \end{array} \right.	\label{eq:clipping} \\
    \kf(n) &= \gamma \kf(n-1) + (1-\gamma){\kappa}_\mathrm{t}(n), \label{eq:freqsmooth}
\end{eqnarray}
\end{subequations}
where 0\;$\ll$\;$\lambdaf$\;$<$\;1 is the forgetting factor, $f(n)$ is the output of the all-pole section, $\kf(n)$ is the estimated parameter ($\wfh(n)=\cos^{-1}\kf(n)$), and $\gamma$ is the smoothing factor. \Eref{eq:latticeinit} sets the initial condition, \mbox{\eref{eq:latticec}--\eref{eq:latticek}} form the frequency estimator, and \eref{eq:clipping} is used to limit ${\kappa}_\mathrm{t}$ in the range of $[-1,1]$ to guarantee stability. \eref{eq:freqsmooth} is used to further smooth $\kf(n)$. For simplicity in notation, in the rest of this paper, $\kf$ is short for $\kf(n)$ and $\wfh$ is short for $\wfh(n)$.

The parameters $\alphaf$ and $\lambdaf$ control the speed and accuracy of frequency estimation. It is advantageous to use time-varying values for $\alphaf$ and $\lambdaf$ due to several reasons. In initial convergence, if the notch is too narrow ($\alphaf$ very close to 1), the ANF may not sense the presence of the input sinusoid, which in turn leads to a very slow initial convergence or even not converging to the correct frequency estimate. Similarly, an initial value of $\lambdaf$ very close to 1, significantly slows down the initial adaptation. On the other hand, smaller values of $\alphaf$ and $\lambdaf$ increase the steady-state error. A solution is to start the algorithm with smaller values of $\alphaf$ and $\lambdaf$ to reach a fast convergence, and after that gradually increase their values to obtain more accurate frequency estimation. For this purpose, $\alphaf$ and $\lambdaf$ are updated in each iteration as
\begin{subequations}
\begin{eqnarray}
    \label{eq:alphalambda}
    \alphaf(n) = \alpha_{\mathrm{st}} \alphaf(n-1) + (1-\alpha_{\mathrm{st}})\alpha_{\infty} \label{eq:alphaf},\\
    \lambdaf(n) = \lambda_{\mathrm{st}} \lambdaf(n-1) + (1-\lambda_{\mathrm{st}})\lambda_{\infty}, \label{eq:lambdaf}
\end{eqnarray}
\end{subequations}
where $\alpha_{\infty}$ determines the asymptotic notch bandwidth and $\alpha_{\mathrm{st}}$ sets the rate of change from the initial value $\alphaf(0)=\alpha_0$ to the asymptotic value $\alpha_{\infty}$. Similarly, $\lambda_{\infty}$ determines the asymptotic forgetting factor and $\lambda_{\mathrm{st}}$ sets the rate of change from initial value  $\lambdaf(0)=\lambda_0$ to the asymptotic value $\lambda_{\infty}$. Detailed discussion on choosing proper values for the parameters are presented in \sref{sec:parameters}

\subsection{Harmonic Estimation} \label{sec:harmonics}
Having estimated $\kf$, the algorithm proceeds to estimate the harmonic components. Harmonic estimation comprises two sub-stages. First, a series of harmonic sinusoids with fundamental frequency $\wfh$ are generated. Subsequently, the amplitudes and the phases of the generated harmonics are estimated to match their corresponding components in the interference.  Harmonic generation is explained in \sref{sec:harmonicgen}, and amplitude/phase estimation is described in \sref{sec:ampphaseest}.

\subsubsection{Harmonic Signal Generation} \label{sec:harmonicgen}
The harmonic sinusoids are generated through using discrete-time oscillators, which require less computation compared with the Taylor expansion method \cite{turner_recursive_2003}.
Among different oscillator structures, a digital waveguide oscillator is chosen (\fref{fig:oscillator}). This structure provides orthogonal outputs, which are exploited to simplify the next stage RLS algorithm. More importantly, the oscillator output frequency can be directly controlled by $\cos k\wfh$, where $k\wfh$ is the oscillation frequency. This enables the output of the frequency estimator $\kf$ to be directly employed for harmonic generation, thus avoiding the calculation of computationally expensive trigonometric functions. To further reduce the complexity, the frequency estimates of higher harmonics are obtained through the recurrence formulation in \eref{eq:recurrence}, which also avoid trigonometric function calculation.
For each harmonic $k$, the frequency control parameter of the oscillator is denoted as $\kappa_k=\cos k\wfh$, and is recursively calculated through
\begin{subequations}
\label{eq:recurrence}
\begin{equation}
		\kappa_k = 2\kappa_1\kappa_{k-1} - \kappa_{k-2},\quad\hbox{for } k=2,3,\cdots, M',
\end{equation}
where
\begin{equation}
	\kappa_0 = 1, \kappa_1 = \kf = \cos\wfh.
\end{equation}
\end{subequations}
The calculated parameter $\kappa_k$ is used to set the oscillation frequency of the oscillator.

\Fref{fig:oscillator} shows the signal flow graph of the digital waveguide oscillator, which is represented by \eref{eq:osc}.
\begin{subequations}
\label{eq:oscillator}
 \begin{eqnarray}
	 \left[\begin{array}{c}
	 	 u_k(n) \\
	 	 u'_k(n)
 	 \end{array} \right]
	=
	\left[ \begin{array}{cc}
		\kappa_k & \kappa_k-1 \\
		\kappa_k+1 & \kappa_k
	 \end{array} \right]
 	 \left[ \begin{array}{c}
	 	 u_k(n-1) \\
	 	 u'_k(n-1)
 	 \end{array} \right],
	\label{eq:osc} \\
 	 G = 1.5 - \left( u_k(n)^2 - \frac{\kappa_k-1}{\kappa_k+1}{u'_k}(n)^2 \right), \label{eq:oscG} \\
 	 u_k(n) = G u_k(n), \quad u'_k(n) = G u'_k(n)\label{eq:oscU}.
 \end{eqnarray}
 \end{subequations}
 Here, $u(n)$ and $u'(n)$ are state variables serving as sinusoidal outputs. The values of $u_k(0)$ and $u'_k(0)$ determine the initial phase and amplitude, which are arbitrarily chosen. \eref{eq:oscG} and \eref{eq:oscU} are used to apply gain control for stabilizing oscillation amplitude in dynamic frequency operation. The output of \eref{eq:oscillator} can be generally expressed as
 \begin{eqnarray}
 	\label{eq:oscV}
 	\eqalign{
 	u_k(n) = v_k\sin(k \wfh n + \psi_k),  \\
 	u'_k(n) = v'_k\cos(k \wfh n + \psi_k),
    }
 \end{eqnarray}
 where $v_k$ and $v'_k$ are the amplitudes of the generated sinusoids, and $\psi_k$ is the initial phase shift. The values of $\psi_k$s do not influence any further derivations and are neglected for simplicity.

\subsubsection{Amplitude and Phase Estimation} \label{sec:ampphaseest}
The amplitudes and phases (i.e.~$\akh$ and $\phikh$) of the generated harmonics are not necessarily the same with their corresponding power line interference components in \eref{eq:pli}; thus, an additional step is required to estimate them. The estimate of the $k^\mathrm{th}$ harmonic, $\hat{h}_k(n)$, can be obtained via \eref{eq:pli} by substituting  $a_k$ and $\phi_k$ with their estimates that gives
\begin{subequations}
\begin{eqnarray}
    \hat{h}_k(n) &= \akh\sin(k\wfh n+\phikh) \label{eq:sin} \\
 	&= \hat{b}'_k\sin(k\wfh n)
    + \hat{c}'_k\cos(k\wfh n). \label{eq:sincos}
\end{eqnarray}
where
\begin{eqnarray}
 	\hat{b}'_k = \hat{a}_k\cos\phikh,	&\nonumber  \quad \mbox{and} \quad
 	\hat{c}'_k = \hat{a}_k\sin\phikh. &\nonumber	
\end{eqnarray}
\end{subequations}
Here, instead of directly adapting $\akh$ and $\phikh$ in \eref{eq:sin} we can equivalently adapt $\hat{b}'_k$ and $\hat{c}'_k$ in \eref{eq:sincos} to obtain $\hat{h}_k(n)$. This transformation converts the non-convex search space in $\hat{a}_k$-$\hat{\phi}_k$ coordinates into a convex search space in rectangular coordinates. Using \eref{eq:oscV} and \eref{eq:sincos}, $\hat{h}_k(n)$ can be written as
\begin{eqnarray}
    \label{eq:harmest}
	\hat{h}_k(n) = 	\hat{b}_k u_k(n) + 	\hat{c}_k u'_k(n).
\end{eqnarray}
Here, $\hat{b}_k$ and $\hat{c}_k$ are defined as $\hat{b}'_k/v_k$ and $\hat{c}'_k/v'_k$, where $v_k$ and $v_k'$ merely scale the adaptive coefficients and do not affect the estimation performance. For each harmonic $k$, $\hat{b}_k$ and $\hat{c}_k$ are adapted by minimizing the exponentially weighted squared error between $\hat{h}_k(n)$ and $x(n)$. This is done by applying the simplified RLS algorithm, where $u_k(n)$ and $u'_k(n)$ serve as the input to an adaptive linear combiner (\fref{fig:combiner}). The following update equations are used to adapt $\hat{b}_k$ and $\hat{c}_k$.
\begin{eqnarray}
	\eqalign{
    r_{1,k}(-1) = r_{1,k}(-1) = \hat{b}_k(-1) = \hat{c}_k(-1) =0, \\
	r_{1,k}(n) = \lambdaa r_{1,k}(n-1) + u_k(n)^2 ,	\\
	r_{4,k}(n) = \lambdaa r_{4,k}(n-1) +  u'_k(n)^2 ,	\\
	\hat{b}_k(n) = \hat{b}_k(n-1) + u_k(n)e_k(n) / r_{1,k}(n) , \\
	\hat{c}_k(n) = \hat{c}_k(n-1) + u'_k(n)e_k(n) / r_{4,k}(n) ,	
    }
\end{eqnarray}
where $e_k(n) = x(n) - \hat{h}_k(n)$ is the instantaneous error (\fref{fig:combiner}), and $0\ll\lambdaa<1$ is the forgetting factor.
A detailed description of the simplified RLS algorithm is described in~\ref{sec:apxRLS}.

In each iteration, the most recent estimates $\hat{b}_k(n)$ and $\hat{c}_k(n)$ are used to obtain $\hat{h}_k(n)$ through \eref{eq:harmest}. The interference-free neural signal is then obtained by
\begin{eqnarray}
    \eqalign{
	 \hat{s}(n) = x(n) - \sum\limits_{k=1}^{M'} \hat{h}_k(n)}.
\end{eqnarray}

\begin{algorithm}

\small
\caption{Proposed Algorithm} \label{algo:APIC}
\DontPrintSemicolon
\KwIn{$x$}
\KwOut{$\hat{s}$ }
\textbf{Constants:} \\
    $\fs$, $M'$, $N$, $\alpha_0$, $\alpha_{\mathrm{st}}$, $\alpha_{\infty}$, $\lambda_0$, $\lambda_{\mathrm{st}}$, $\lambda_{\infty}$, $\lambdaa$, $\gamma$\;
    $H(\cdot) \gets$ 40--70~Hz IIR filter\;
\textbf{Initialization:}\\
	$\kappa_{0} \gets 1$,  $\kf \gets 0$\;
	$f_{-2} \gets f_{-1} \gets 0$\;
	$  c, d > 0$\;
    $u_k, u'_k > 0$\;
    $ r_{1,k}, r_{4,k} > 0$\;
    $\hat{b}_k \gets \hat{c}_k \gets 0$\;
    $\alphaf \gets \alpha_0$, $\lambdaf \gets \lambda_0$\;
\textbf{Recursion:}\\
\For{$n \gets 1$ \textbf{to} $N$}{
	\emph{Bandpass filtering:}\\
	\enskip $x_f \gets H(x(n))$\;
	\emph{Frequency Estimation:}\\
	\enskip $f_n \gets x_f + \kf(1+\alphaf)f_{n-1} - \alphaf f_{n-2}$\;
	\enskip $c \gets \lambdaf c + f_{n-1}(f_n+f_{n-2})$\;
	\enskip $d \gets \lambdaf d + 2 f^2_{n-1}$\;
	\enskip ${\kappa}_\mathrm{t} \gets c/d$\;
	\enskip \lIf{${\kappa}_\mathrm{t} > 1$ }{${\kappa}_\mathrm{t} \gets 1$}
  	\enskip \lElseIf{${\kappa}_\mathrm{t} < -1$ }{${\kappa}_\mathrm{t} \gets -1$}
	\enskip $\kf \gets \gamma \kf + (1-\gamma){\kappa}_\mathrm{t}$\;
	\enskip $\alphaf \gets \alpha_{\mathrm{st}}\alphaf + (1-\alpha_{\mathrm{st}})\alpha_{\infty}$\;
	\enskip $\lambdaf \gets \lambda_{\mathrm{st}} \lambdaf + (1-\lambda_{\mathrm{st}})\lambda_{\infty}$ \;
	\emph{Removing Harmonics:}\\
	\enskip $\kappa_{1} \gets \kf$\;
	\enskip $e \gets x(n)$\;
    \enskip \For{$k \gets 1$ \textbf{to} $M'$}{
	\emph{Discrete Oscillator:}\\
	\enskip $s_1 \gets \kappa_k(u_k + u'_k)$\;
	\enskip $s_2 \gets u_k$\;
	\enskip $u_k \gets s_1 - u'_k$\;
	\enskip $u'_k \gets s_1 + s_2$\;
	\enskip $G \gets 1.5 - [u_k^{2} - u'^{2}_k(\kappa_k-1)/(\kappa_k+1)]$\;
	\enskip \lIf {$G <0$}{$G \gets 1$} 
	\enskip $u_k = G u_k$, $u'_k = G u'_k$\;
	\emph{Amplitude/Phase Estimation:}\\
	\enskip	$h_k \gets (\hat{b}_k u_k + \hat{c}_k u'_k)$\;
    \enskip  $e \gets e - h_k$\;
	\enskip $r_{1,k} \gets \lambdaa r_{1,k} + u^2_k$\;
	\enskip $r_{4,k} \gets \lambdaa r_{4,k} + u'^2_k$\;
	\enskip	$\hat{b}_k \gets \hat{b}_k  + e\cdot u_k /r_{1,k}$\;
	\enskip	$\hat{c}_k \gets \hat{c}_k  + e\cdot u'_k /r_{4,k}$\;	
	\emph{Harmonic Frequency Calculation:}\\
	\enskip $\kappa_{k+1} \gets 2\kf\kappa_{k} - \kappa_{k-1}$\;
	}
	\enskip $\hat{s}(n) \gets e$\;}
\end{algorithm}

\begin{table}
\caption{\label{tab:algvar}List of symbols and parameters}
\centering
\begin{tabular}{@{}c@{ }l@{}}
\toprule
Symbol & Explanation\\
\midrule
$N$ & Number of samples\\

$\fs$ & Sampling rate (Hz)\\

$M'$ & Number of harmonics to remove\\

$B_0$ & Initial notch bandwidth of the frequency estimator (Hz)\\

$B_{\infty}$ & Asymptotic\,notch\,bandwidth\,of\,the\,frequency\,estimator\;(Hz)\\

$B_{\mathrm{st}}$ & Settling time from $B_0$ to $B_{\infty}$ (s)\\

$\alphaf$ & Pole radii of the adaptive notch filter (ANF)\\

$\alpha_0$ & Initial pole radii of the ANF\\

$\alpha_{\infty}$ & Asymptotic pole radii of the ANF\\

$\alpha_\mathrm{st}$ & Rate of change from $\alpha_0$ to $\alpha_{\infty}$\\

$P_0$ & Initial settling time of the frequency estimator (s)\\

$P_{\infty}$ & Asymptotic settling time of the frequency estimator (s)\\

$P_{st}$ & Settling time from $P_0$ to $P_{\infty}$ (s)\\

$\lambdaf$ & Forgetting factor of the frequency estimator\\

$\lambda_0$ & Initial forgetting factor of the frequency estimator\\

$\lambda_{\infty}$ & Asymptotic forgetting factor of the frequency estimator\\

$\lambda_\mathrm{st}$ & Rate of change from $\lambda_0$ to $\lambda_{\infty}$ \\

$\gamma$ & Smoothing parameter of the frequency estimator \\

$\gamma'$ & Cut-off frequency of the smoothing filter; set at 90~Hz \\

$W$ & Settling time of amplitude/phase estimator (s) \\

$\lambdaa$ & Forgetting factor of the amplitude/phase estimator\\

$H(\cdot)$ & 40--70~Hz \supsc{4}{th} order IIR bandpass filter \\
\bottomrule
\end{tabular}
\end{table}

\subsection{Algorithm Implementation}
The algorithm is implemented in software as well as an ASIC.
The pseudocode of the algorithm is presented in algorithm~\ref{algo:APIC} with the MATLAB source code available online at \cite{MATLABCode}. An explanatory list of the symbols and parameters is shown in table~\ref{tab:algvar}, and the proper parameter values can be obtained through the guidelines in \sref{sec:parameters}.

The ASIC was fabricated in a 65-nm CMOS process, and consumes 0.11~mm$^2$ of silicon area. It was tested against a reference model, where its robust and real-time operation was experimentally verified. For validating the chip output, we used a full-precision MATLAB implementation of the algorithm with the same structure and parameter values used in the chip design. This implementation is referred to as `reference model' in the rest of this paper.
Further discussion of hardware implementation and testing results are presented in \ref{sec:apxHardware}.

\subsection{Parameter Setting} \label{sec:parameters}
The performance of the algorithm is mainly controlled by three basic parameters including notch filter pole radii ($\alphaf$), frequency estimator's forgetting factor ($\lambdaf$) and amplitude/phase estimator's forgetting factor ($\lambdaa$).
Since, the proper values of these parameters depend on the signal sampling rate ($\fs$), the parameter adjustment become less intuitive. To alleviate this issue, we have chosen other representative characteristics such as notch bandwidth (related to the pole radii) and settling time (related to the forgetting factors) which can be alternatively used for parameter adjustment. The alternative parameters are displayed in \eref{equ:paramalt}.
\begin{subequations}
\label{equ:paramalt}
\begin{eqnarray}
\label{equ:paramset}
\eqalign{
\mathcal{A}_1 = \{\alpha_\mathrm{st},\lambda_0, \lambda_{\infty},\lambda_\mathrm{st} ,\lambdaa\}, \\
\mathcal{B}_1 = \{ B_\mathrm{st}, P_0, P_{\infty},P_\mathrm{st}, W \}, \\
\mathcal{A}_2 = \{ \alpha_0 , \alpha_{\infty}, \gamma'\}, \\
\mathcal{B}_2 = \{B_0 , B_{\infty}, \gamma/2 \},
}
\end{eqnarray}
\begin{eqnarray}
\label{equ:parammap}
\eqalign {
\mathcal{A}_1 = \exp{{\frac{\ln(0.05)}{\mathcal{B}_1\fs+1}}} \label{equ:lamset}, \\
\mathcal{A}_2 = \frac{1-\tan{(\pi\mathcal{B}_2/\fs)}}{1+\tan{(\pi \mathcal{B}_2/\fs)}}.
}
\end{eqnarray}
\end{subequations}
Here, $\mathcal{B}_1$ and $\mathcal{B}_2$ contain the alternative parameters which are independent of the sampling rate and have intuitive units. The actual parameters, defined in $\mathcal{A}_1$ and $\mathcal{A}_2$, can be obtained through \eref{equ:parammap}. It should be noted that improper parameter setting may lead to inadequate removal of the interference. Some guidelines on the proper adjustment of the parameters are discussed as follows.

The notch bandwidth of the frequency estimator affects both the tracking speed and the estimation bias. A wide notch allows faster tracking of the frequency at the expense of an increased estimation bias and variance. On the other hand, a narrow notch leads to a more accurate frequency estimate, but it causes very slow frequency adaptation if the desired sinusoidal component falls out of the notch bandwidth.
To address this trade-off, the notch bandwidth is initially widened to allow fast initial convergence and then gradually narrowed down to achieve a lower steady-state error (described in \eref{eq:alphaf}).
In the alternative form, $B_0$ is associated with $\alpha_0$ and controls the initial notch bandwidth. Larger values of $B_0$ are preferred (e.g.~tens of~Hz) to achieve a faster initial convergence. Similarly, $B_{\infty}$ is associated with $\alpha_{\infty}$ and controls the asymptotic notch bandwidth. Small values of $B_{\infty}$ are preferred (e.g.~tenths of~Hz) to achieve more accurate estimation of the frequency. $B_{\mathrm{st}}$ controls the rate of transition between initial notch bandwidth $B_0$ and the asymptotic notch bandwidth $B_{\infty}$, and indicates the time in seconds, in which $\alphaf$ reaches $0.95 \alpha_{\infty}$ in \eref{eq:alphaf}. When the algorithm is used to remove a large number of harmonic components, $B_{\infty}$ should be set small enough to minimize the bias in the frequency estimates of higher harmonics. For example, if it is desired to remove harmonics up to \supsc{100}{th} order, setting $B_{\infty}=0.001$ would be an adequate choice. In this case, although small values of $B_{\infty}$ lead to slow frequency adaptation, it would not be problematic, since in practice, the drifts in the power line frequency are usually slow and the algorithm can still reasonably track the variations.

The forgetting factor of the frequency estimator $\lambdaf$ is initially small to achieve a fast convergence and is gradually increased to achieve a more accurate estimate (described in \eref{eq:lambdaf}). In the alternative form, $P_0$ is associated with $\lambda_0$ and controls the initial settling time of the frequency estimation algorithm. Smaller values of $P_0$ are preferred (e.g.~tenths of seconds) to achieve a faster initial convergence. Similarly, $P_{\infty}$ is associated with $\lambda_{\infty}$ and controls the asymptotic settling time of the frequency estimation algorithm. Considering the fact that the power line frequency drifts are slow, larger values of $P_{\infty}$ are preferred (e.g.~a few seconds) to obtain a more accurate estimation of the power line frequency. $P_{\mathrm{st}}$ controls how fast the settling time changes from the initial value of $P_0$ to its final value of $P_{\infty}$ and indicates the time in seconds, in which $\lambdaf$ reaches $0.95 \lambda_{\infty}$ in \eref{eq:lambdaf}. This transition time should be set large enough (e.g.~a few seconds depending on the notch bandwidth) to allow global convergence.

The settling time of the amplitude/phase estimator ($W$) controls how fast it responds to the fluctuations in the amplitudes and phases of the harmonics. In the alternative form, $W$ is associated with $\lambdaa$, and indicates the time in which the estimates of amplitude and phase reach 95\% of their asymptotic values. The interference frequency bands (e.g.~near 50/60~Hz and multiples) contain both the interference components as well as useful neural signals which should be preserved. For this purpose, $W$ should be selected reasonably large to obtain an accurate estimation of the interference, thus avoiding the excessive removal of neural signals, while small enough to allow tracking of the interference amplitude fluctuations. Depending on the recording environment and subject movements, $W$ may be selected from a few tenths of seconds to a few seconds.
A recommended set of parameter values are suggested in table~\ref{tab:properparam} which could be initially used for further tuning.

\begin{table}[h]
\caption{\label{tab:properparam} Recommended values of parameters}
\centering
\begin{tabular}[t]{@{\quad }l@{ }>{\scriptsize}ld{4.4}@{\ }}
\toprule
\multicolumn{2}{c}{Parameter} &\multicolumn{1}{@{}c}{Recommended Range}\\
\midrule
$B_0$ & (Hz) & 10 - 50  \\
$B_{\infty}$ & (Hz) & 0.01 - 0.1 \\
$B_\mathrm{st}$ & (s) & 0.5 - 10 \\
$P_0$ & (s) & 0.01 - 0.5\\
$P_{\infty}$ & (s) & 1 - 5 \\
$P_\mathrm{st}$ & (s) &1 - 10 \\
$W$ & (s) &0.5 - 5\\
\bottomrule
\end{tabular}
\end{table}

\begin{figure}[t]
    \centering
    \includegraphics[page=1]{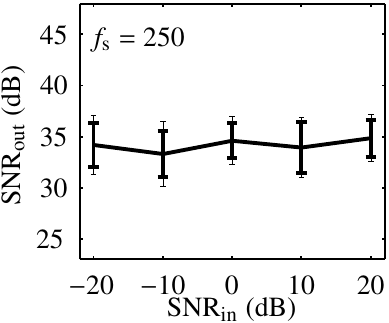} \quad
    \includegraphics[page=2]{figure4_5_12}	\\
    \vspace{2mm}
    \includegraphics[page=3]{figure4_5_12} \quad
    \includegraphics[page=4]{figure4_5_12}
    \caption{ \SNRout vs. \SNRin. Each figure is obtained at a different sampling rate. The horizontal plots indicate the mean, the thick bars show the standard deviation and the thin bars indicate minimum and maximum \SNRout values over 50 runs on real ECoG signals with synthetic interference containing 3 harmonics at 61~Hz, 122~Hz, and 183~Hz (2 harmonics for $\fs$=250~Hz).  Consistent high values of \SNRout indicate the robust operation of the algorithm with regard to different \SNRin and sampling rates. Parameter setting: \{$B_0 = 50$, $B_{\mathrm{st}} = 1$, $B_{\infty} = 0.1$, $P_0 = 0.1$, $P_{st} = 1$, $P_{\infty}=2$, $W = 2$\}.} \label{fig:SNRin}
\end{figure}

\begin{figure}[t]
	\centering
	\includegraphics[page=5]{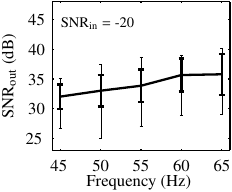}  \quad
	\includegraphics[page=6]{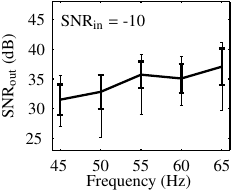}	\\
		\vspace{2mm}	
	\includegraphics[page=7]{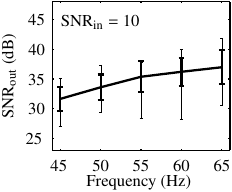} \quad
	\includegraphics[page=8]{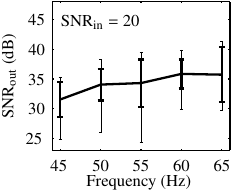}
	\caption{\SNRout vs. power line frequency. The horizontal plots indicate the means, the thick bars show the standard deviation and the thin bars indicate minimum and maximum \SNRout values over 50 runs on real ECoG signals with synthetic interference containing 3 harmonics.
Consistent high values of \SNRout are achieved in the wide range of power line frequencies and sampling rates.
The reason for the slight increase of mean {\SNRout} with frequency is mainly due to the \mbox{$1/f$} PSD of neural signals. With a fixed \SNRin, at higher frequencies, the power of neural signals are less, leading to a more accurate estimation of the interference (neural signals are seen as noise to the interference estimation algorithm), hence resulting in a better cancellation and slightly higher {\SNRout} compared with the lower frequencies.
Parameter setting is the same as that of \fref{fig:SNRin}.}
	\label{fig:powerfreq}
	\end{figure}

\section{Results} \label{sec:simulations}
Extensive simulations are carried out to quantitatively evaluate the performance of the proposed algorithm under various signal and parameters conditions. The algorithm performance is also compared with other popular interference removal methods. Furthermore, the algorithm is tested on extracellular, electrocorticography (ECoG) and electroencephalography (EEG) recordings to illustrate its performance on real neural data.
The results of the performance evaluation using synthetic data are described in \sref{sec:synthsim}, the performance comparison results are reported in \sref{sec:comparison}, and the results on real data are presented in \sref{sec:realsim}. In case the reader wishes to reproduce the paper's results, the parameter setting in each simulation is provided.

\subsection{Performance Evaluation on Synthetic Data} \label{sec:synthsim}
Synthetic data are used to quantitatively evaluate the important characteristics of the proposed algorithm under various signal conditions.
Each test sequence was synthesized by adding a synthetic interference containing 3 harmonics, to a random portion of real ECoG and extracellular recordings that were recorded in a controlled condition with negligible amount of power line interference. The frequency and power of the interference components are specified in each simulation.

In the rest of this paper, \SNRin and \SNRout are used to denote the SNRs of the algorithm input and output signals, i.e.~$x(n)$ and $\hat{s}(n)$, respectively. It should be noted that, \SNRout values are calculated after the algorithm reaches its steady-state, unless otherwise stated.

\subsubsection{Sensitivity to \SNRin}
The variations in the power of the picked-up interference are usually significant,
 leading to different \SNRin values from as low as $-20$~dB (severe interference), to as high as 30~dB (negligible interference). To ensure proper interference cancellation, the algorithm is desired to work reliably under various \SNRin conditions. To evaluate this aspect, we generated synthetic sequences whose \SNRin ranged from $-20$~dB to 20~dB. For each \SNRin value, 50 sequences were generated, the algorithm was applied to cancel the interference, and the resultant \SNRout{}s were recorded. In addition the simulation was repeated with different sampling rates for reliability resting.

\Fref{fig:SNRin} shows the mean, variance, minimum and maximum of the resultant \SNRout for each \SNRin condition and sampling rate. It can be seen that, consistent high values of \SNRout are observed in all the conditions, indicating that the performance of the algorithm is highly insensitive to \SNRin.

	\begin{figure*}[t]
    	\centering
        \includegraphics{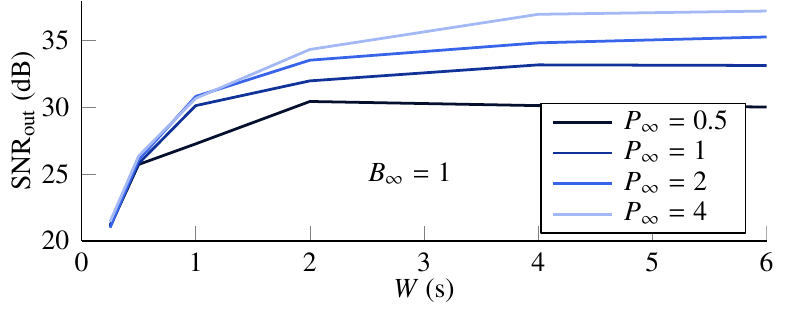}
        \hfil
        \includegraphics{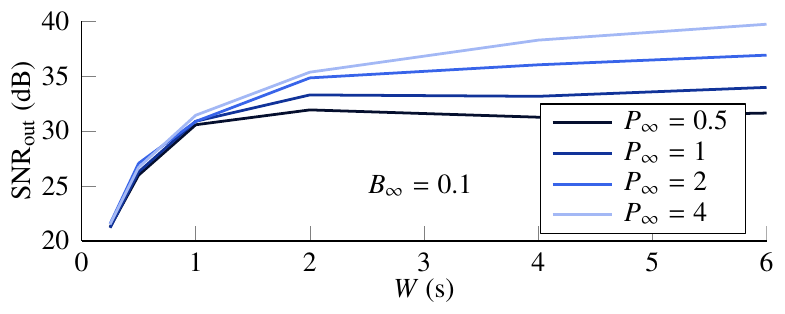}
    	\\
        \includegraphics{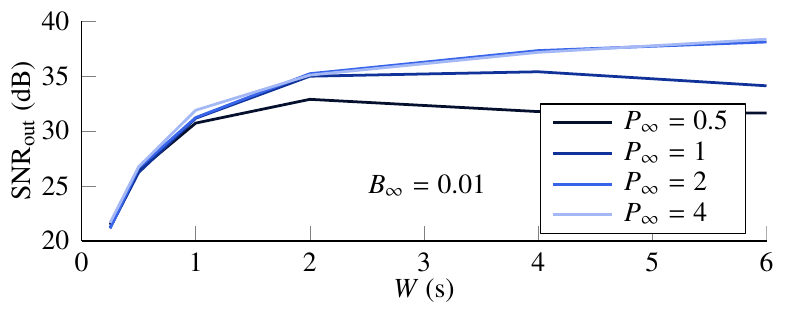}
        \hfil
        \includegraphics{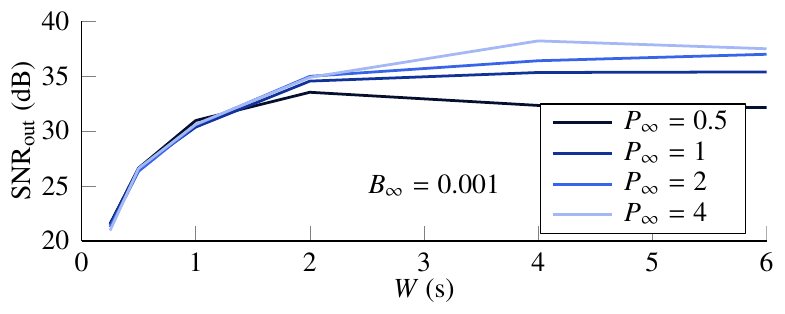}
    	\caption{\label{fig:snr-time} Trade-off between amplitude settling time and \SNRout. The plots display the \SNRout versus amplitude settling time $W$, for different $P_{\infty}$ and $B_{\infty}$. \SNRin is set to ~0~dB for all the cases. The results show that high \SNRout values ($>$~30~dB) can be achieved along with a reasonably fast settling time ($<1$~s at $W=1$). Parameter setting: \{$\fs=1$~kHz, $B_0=50$, $B_{\mathrm{st}}=1$, $P_0=0.1$, $P_{st}=1$\} }
	\end{figure*}

\begin{figure*}[t]
\captionsetup[subfigure]{labelformat=empty}
\centering
\subfloat[][\footnotesize (a-1) \supsc{1}{st} harmonic]{\includegraphics{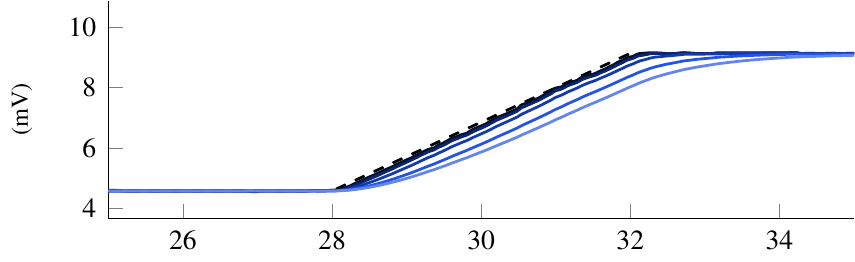}} \hfill
\subfloat[][\footnotesize (b-1) \supsc{1}{st} harmonic]{\includegraphics{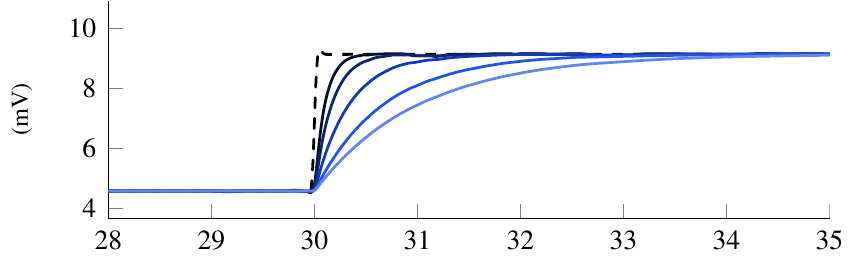}} \\
\vspace{-3mm}
\subfloat[][\footnotesize (a-2) \supsc{2}{nd} harmonic]{\includegraphics{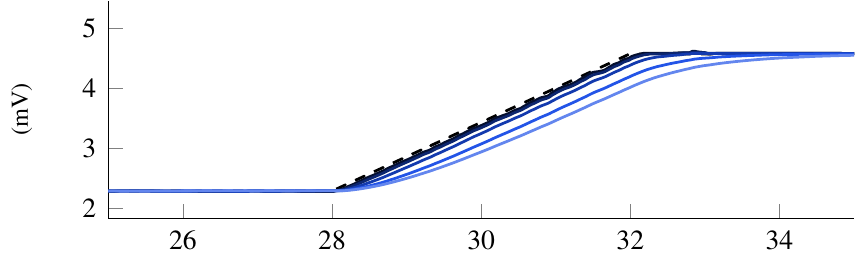}} \hfill
\subfloat[][\footnotesize (b-2) \supsc{2}{nd} harmonic]{\includegraphics{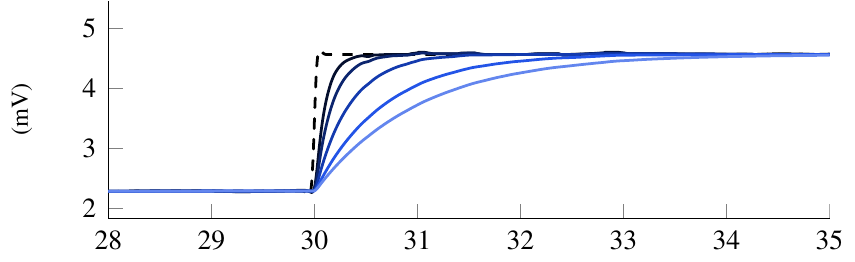}} \\
\vspace{-3mm}
\subfloat[][\footnotesize (a-3) \supsc{3}{rd} harmonic]{\includegraphics{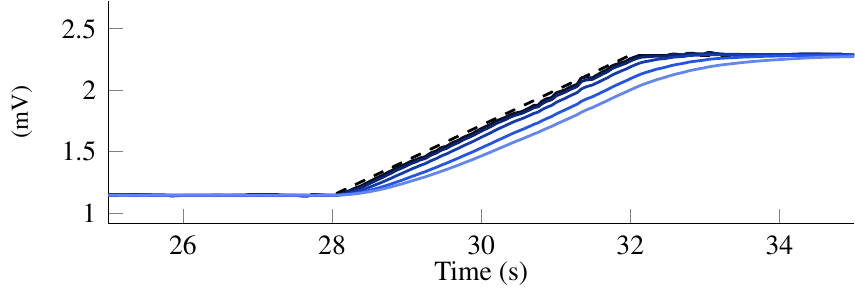}} \hfill
\subfloat[][\footnotesize (b-3) \supsc{3}{rd} harmonic]{\includegraphics{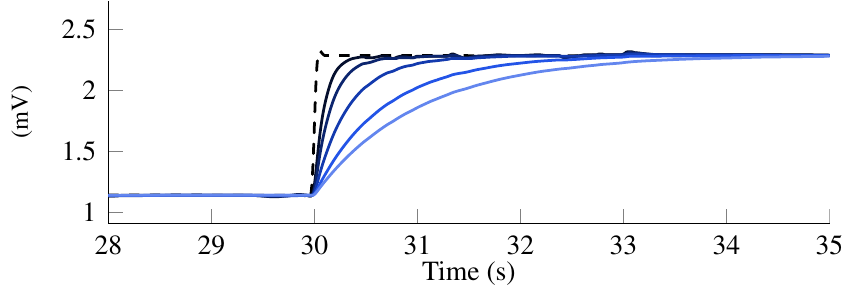}}\\
\includegraphics{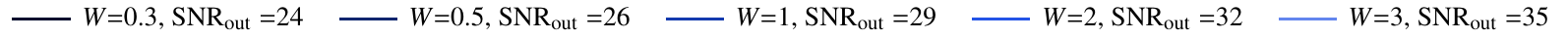}\\
\caption{Amplitude tracking. In (a-1)-(a-3),  the amplitudes of the harmonics were gradually increased to twice their initial values. In (b-1)-(b-3), the amplitudes of the harmonics underwent a step jump. The actual amplitude is displayed by (\protect\plotref{plt:ampltrackdashed}). The algorithm was applied with different values of $W$ which led to different settling times and \SNRout values. The input SNR was set to \SNRin~=~0~dB, and {\SNRout} values were calculated after convergence ($t>35$~s). As can be seen, smaller values of $W$ have led to faster amplitude tracking, however yielded lower \SNRout. On the other hand, larger values of $W$ resulted in a slower amplitude tracking but yielded higher \SNRout. Parameter setting: \{$\fs=1$~kHz,  $B_0=50$, $B_{\mathrm{st}}=1$, $B_{\infty}=0.1$, $P_0=0.1$, $P_{st}=1$, $P_{\infty}=1$\} }
\label{fig:ampconv}
\end{figure*}
	
\subsubsection{Sensitivity to Power Line Frequency}
Since the accurate value of power line frequency is a priori unknown, and may also change over time \cite{dugan_electrical_2003,baggini_handbook_2008}, it is important to test the performance of the algorithm with regard to different power line frequencies. For this purpose, synthetic sequences with fundamental frequencies ranging between 45~Hz to 65~Hz were used as the input to the algorithm, and output SNRs were measured to test the performance. This frequency range covers the worst case power line frequency deviations \cite{dugan_electrical_2003,baggini_handbook_2008}.  As can be seen in \fref{fig:powerfreq}, high values of \SNRout ($>$~30~dB) were consistently achieved for different power line frequencies in all the \SNRin conditions. The results demonstrate the robust operation of the algorithm even in worst case power line frequency deviations. Furthermore, it can be seen that the algorithm can automatically detect the interference at 50~Hz or 60~Hz, and no a priori setting of the nominal power line frequency is required.
\begin{figure*}[t]
\captionsetup[subfigure]{oneside,margin={6mm,0mm}}
	\centering
	\subfloat[][]{\includegraphics{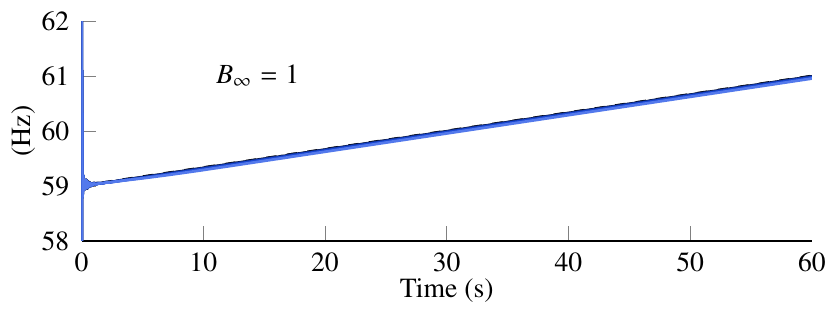}\label{fig:freq1}} \hfill
	\subfloat[][]{\includegraphics{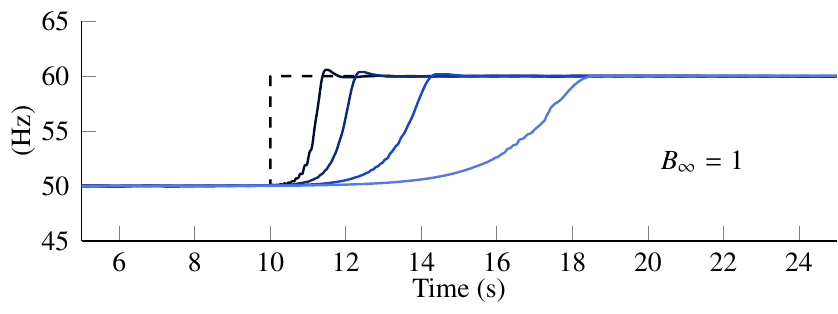}\label{fig:freq2}} \\
    \includegraphics{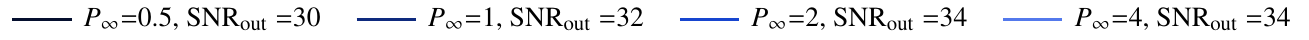}\\

	\subfloat[][]{\includegraphics{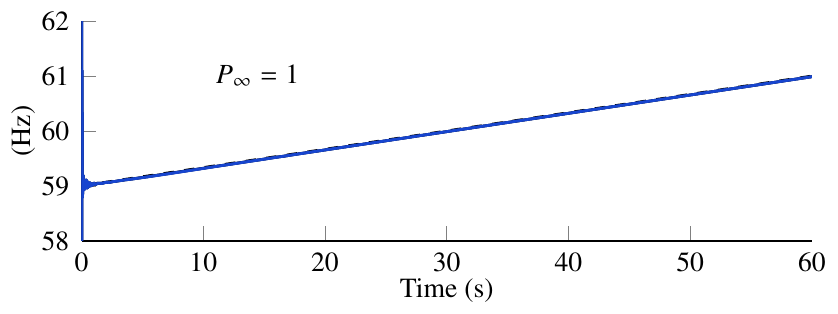}\label{fig:freq3}} \hfill
	\subfloat[][]{\includegraphics{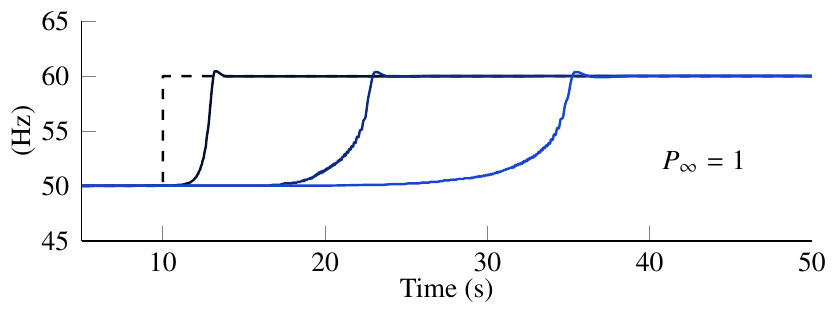}\label{fig:freq4}} \\
    \includegraphics{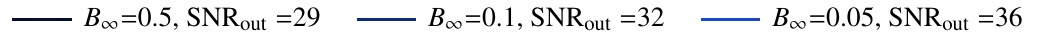}
	\caption{\label{fig:freqconv} Frequency tracking. The actual frequency is shown by (\protect\plotref{plt:ampltrackdashed}). In (a) and (c) the fundamental frequency is swept from 59~Hz to 61~Hz. It can be seen that, the estimated values properly  track the changes of the frequency, with the minimum \SNRout =  26~dB during tracking. In (b) and (d), the fundamental frequency is abruptly changed from 50~Hz to 60~Hz, and the frequency estimates have well track the change. In this simulation, \SNRin = 0~dB and \SNRout values are calculated after convergence (t\;$>$\;20~s in (b) and t\;$>$\;40~s in (d)). In (a) and (b), $B_{\infty}=1$ and $P_{\infty}$ was varied. In (c) and (d), $P_{\infty}=1$ and $B_{\infty}$ was varied. Parameter setting: \{$\fs=1$~kHz,  $B_0=50$, $B_{\mathrm{st}}=1$, $P_0=0.1$, $P_{st}=1$, $W=1$\}}
\end{figure*}

\subsubsection{Trade-off between Settling Time and \SNRout}
As discussed in \sref{sec:parameters}, there is a trade-off between \SNRout and the amplitude settling time ($W$).
To track the abrupt changes in the interference power, fast settling time is desired. Typically, when $W$ is set small to have a fast tracking response, the \SNRout would decrease. On the other hand, when $W$ is set large to achieve a higher \SNRout, then the settling time will increase. It is desirable to achieve a high  \SNRout along with a reasonably fast settling time. \Fref{fig:snr-time} shows the average values of \SNRout versus different settling time values ($W$). It can be seen that high values of \SNRout ($\approx$\,30~dB) can be obtained with a reasonably low settling time (\textless~1~s).

\begin{figure}[t]
	\centering
	\scriptsize \includegraphics{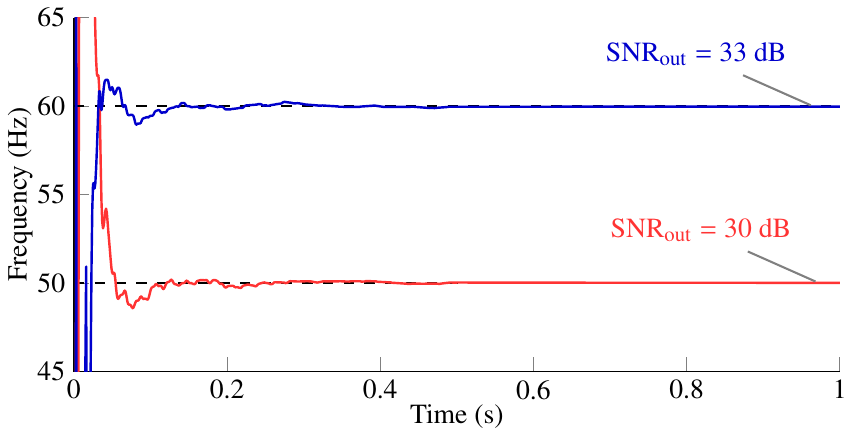}
	\caption{ \label{fig:init-freq} The initial convergence to two interference frequencies at 50~Hz and 60~Hz. A quick initial frequency convergence can be observed. In this simulation \SNRin = 0~dB, and the values of \SNRout were calculated after $t=1$~s. Parameter setting: \{$\fs=1$~kHz,  $B_0=50$, $B_{\infty}=0.05$, $B_{\mathrm{st}}=0.5$, $P_0=0.1$, $P_{\infty}=2$, $P_{st}=0.5$, $W=1$\}}
\end{figure}

\begin{figure}[]
	\centering
	\includegraphics{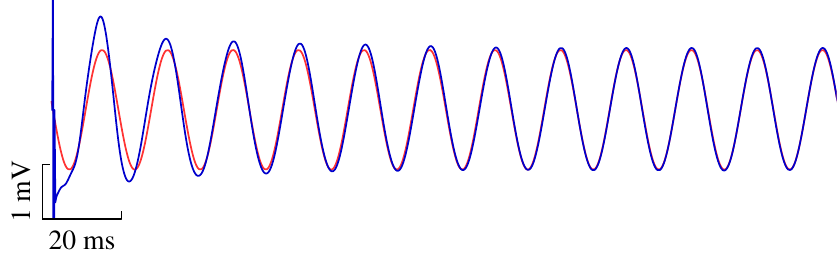}
    \\
   	\includegraphics{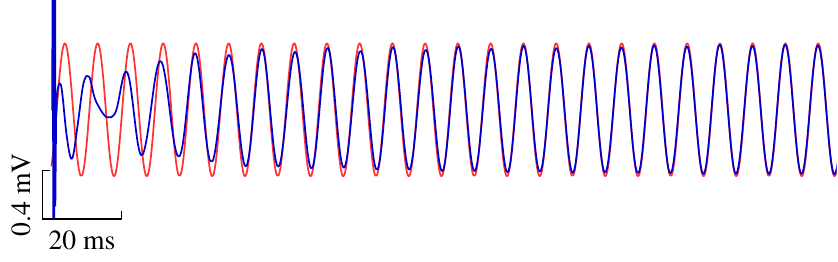}
	\\
	\includegraphics{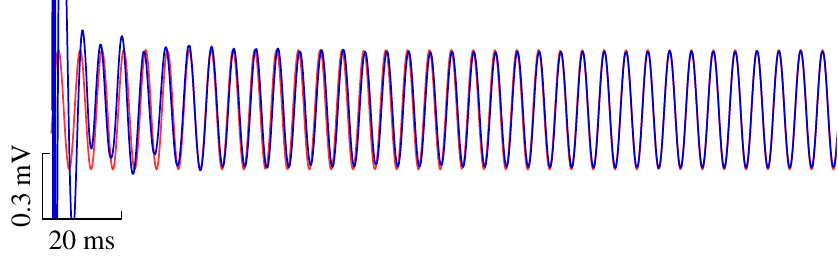}
    \\
   	\includegraphics{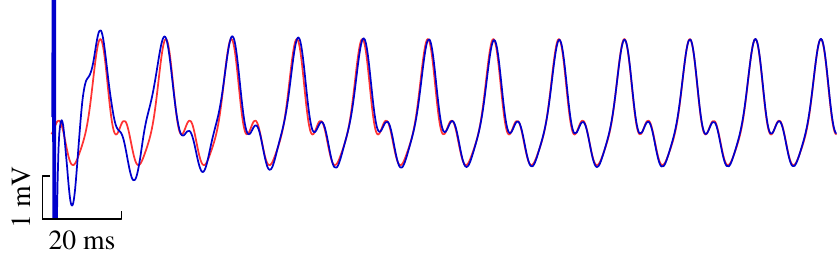}
	\caption{Initial convergence of the estimated harmonics. The method shows a fast adaptation of frequency, phase and amplitude. From top to bottom, $1^\mathrm{st}$, $2^\mathrm{nd}$, $3^\mathrm{rd}$ harmonic and total interference. The plots show actual components (\ref{plt:orgint}) and the estimates (\ref{plt:estint}). In this simulation, \SNRin = 0~dB and \SNRout = 33~dB (for $t>1~s$). Parameter setting is the same as that of \fref{fig:init-freq}. }\label{fig:init-harm}
\end{figure}

\begin{figure*}
	\centering
    \captionsetup[subfigure]{oneside,margin={6mm,0mm}}
	\subfloat[][]{\includegraphics{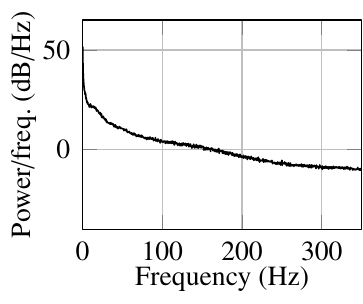}}
	\;
    \subfloat[][]{\includegraphics{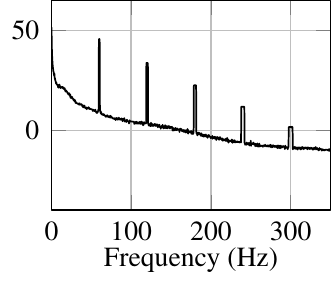}\label{fig:interf}}
    \;
	\subfloat[][]{\includegraphics{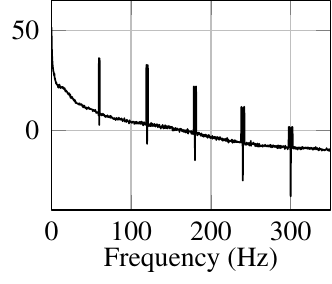} \label{fig:narrownotch}}
	\;
	\subfloat[][]{\includegraphics{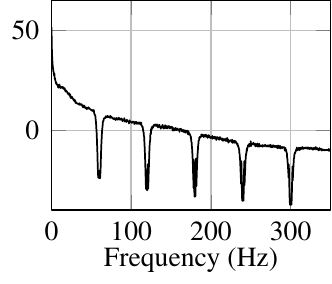}\label{fig:widenotch}}
    \;
    \subfloat[][]{\includegraphics{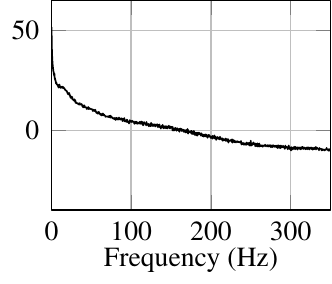}\label{fig:proposed}}
		\caption{The effect of notch filtering on ECoG signal corrupted with power line interference. (a) PSD of the actual ECoG signal. (b) PSD of the synthetically corrupted signal. The interference fundamental frequency is slightly deviated from 60~Hz. (c) PSD after applying narrow-band (1~Hz) IIR notch filters centred at 60~Hz and multiples. (d) PSD after applying wide-band (8~Hz) notch filters. (e) PSD after interference cancellation with the proposed algorithm, where the interference is completely removed while the signal frequency bands are minimally affected. The narrow-band filter fails to adequately remove the interference due to its changing frequency. The wide-band filter distorts the signal spectrum at the rejection bands.}\label{fig:notchfilt}
 \end{figure*}

\subsubsection{Tracking of Amplitude and Frequency Fluctuations}
The drifts of the power line frequency are typically small, while the fluctuations of the harmonics amplitudes can be quite large \cite{dugan_electrical_2003,baggini_handbook_2008}.
In order to effectively reject the power line interference, the algorithm should be able to adequately track the frequency and amplitude variations.

To illustrate the amplitude tracking performance, the harmonic amplitudes were increased to twice their initial values and the algorithm was applied with different settling time values ($W$). \Fref{fig:ampconv} displays the first three interference harmonics, where the they underwent a ramp change and a step change.
It can be seen that, the estimates of the amplitudes properly tracked the actual values.

To illustrate the frequency tracking performance, two simulations were done. In the first simulation, the fundamental frequency of the synthetic harmonics was swept from 59~Hz to 61~Hz. It can be seen in figures~\ref{fig:freq1}~and~\subref*{fig:freq3} that, for all the parameter conditions, the frequency estimates accurately track the actual values. In the second simulation, the fundamental frequency underwent a step change from 50~Hz to 60~Hz. Figures~\ref{fig:freq2}~and~\subref*{fig:freq4} show that, the frequency estimate converges to the actual frequency with different settling times which depend on the parameters $B_{\infty}$ and $P_{\infty}$. It should be noted that, due to the use of time-varying parameters in \eref{eq:alphalambda}, the initial convergence is much faster than is in the operating condition.

\subsubsection{Initial Convergence}
To illustrate the convergence behaviour of the algorithm, two synthetic sequences with interference fundamental frequency of 50~Hz and 60~Hz were used. The interference contained 3 harmonics.
\Fref{fig:init-freq} shows the frequency convergence, where the frequency estimates converged to the actual frequencies (i.e.~ 50~Hz and 60~Hz) in less than 100~ms, while maintaining a high \SNRout. This fast convergence speed is mainly due to adopting time-varying $\alphaf$ and $\lambdaf$. In other words, the initial convergence is controlled by the parameters $B_0$, $B_{\mathrm{st}}$, $P_0$ and $P_{\mathrm{st}}$, whereas the parameters $B_{\infty}, P_{\infty}$ and $W$ determine \SNRout. The convergence of the three estimated harmonics is displayed in \fref{fig:init-harm}, where a quick ($<100$~ms) convergence to actual harmonics is observed.

\begin{figure}[t]
	\centering
	\includegraphics[page=9]{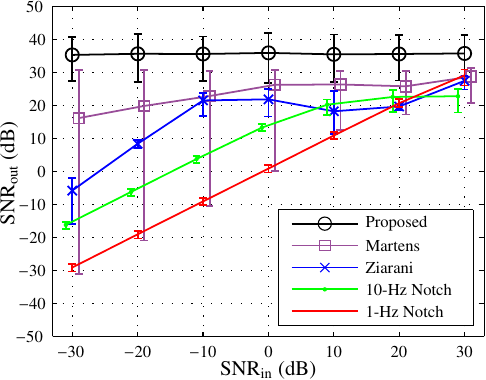}
\caption{Comparison of asymptotic performances of different interference removal methods. The plots represent the average and the error bars indicate the maximum and minimum {\SNRout}. For each \SNRin, the {\SNRout} values are obtained throughout 200 independent runs on synthesized sequences of corrupted ECoG signals. The interference consists of a stationary sinusoid fixed at 59~Hz. The {\SNRout} values are calculated after t\;=\;60~s to ensure the full convergence of the algorithms. When applied on ECoG signals, the proposed algorithm consistently yields high {\SNRout}. The Martens' algorithm performs well in the mean sense; however, large deviations of minimum {\SNRout} imply that it may not always converge. The Ziarani's algorithm is sensitive to the input signal power, thus same parameters cannot be used to achieve an optimal performance for different values of \SNRin. The 10-Hz and 1-Hz notch filters are centred at 60~Hz. The parameter setting is the same as that of \mbox{\fref{fig:MSE_conv}}. \label{fig:SNRin_SNRout_errbars}  }
\vspace{-10pt}
\end{figure}

\begin{figure*}[t]
	\centering
\captionsetup[subfigure]{oneside,margin={6mm,0mm}}
    \subfloat[][]{\includegraphics{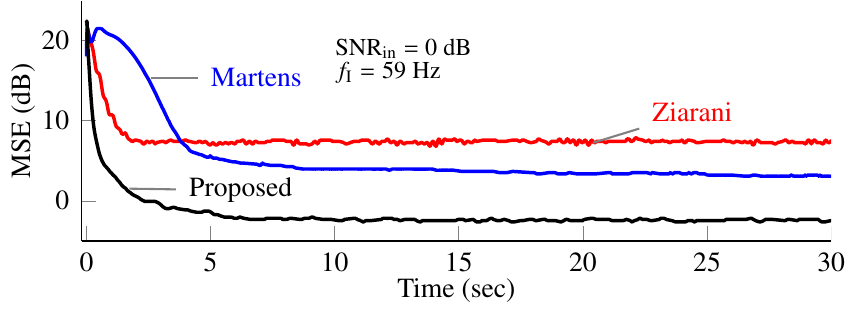}}
    \hfil
    \subfloat[][]{\includegraphics{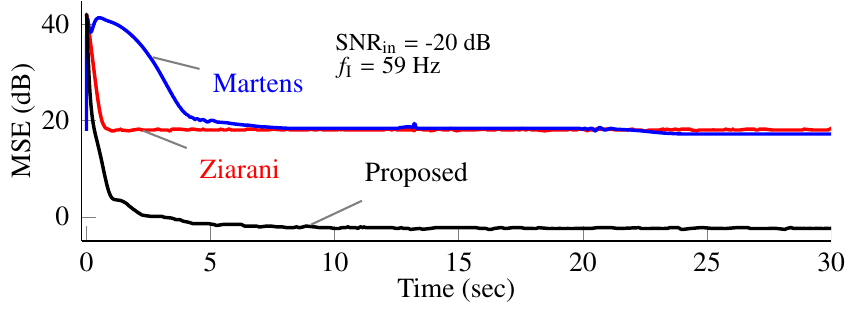}}
    \\
    \subfloat[][]{\includegraphics{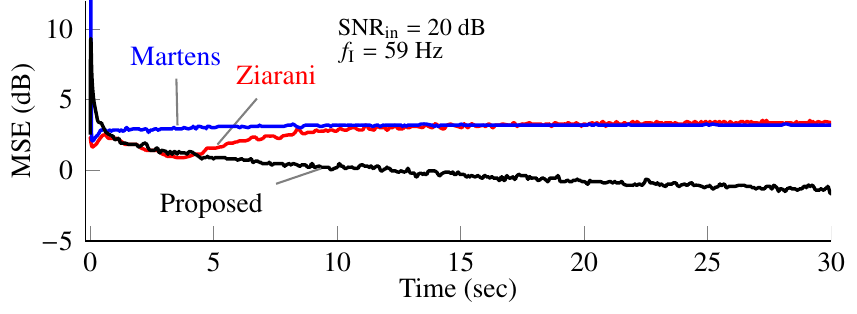}}
    \hfil
    \subfloat[][]{\includegraphics{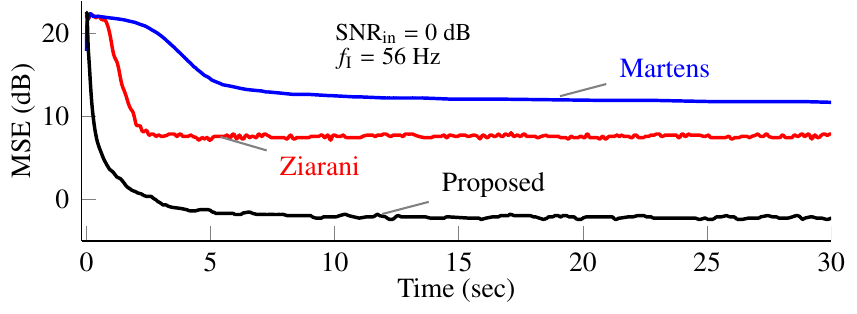}}
	\caption{\label{fig:MSE_conv} Learning curves of the proposed, Ziarani's \cite{ziarani_nonlinear_2002} and Martens'  \cite{martens_improved_2006} algorithms when applied to random signals with $1/f$ PSD superimposed with a single stationary sinusoid at $f_{\mathrm{I}}$~Hz. The mean square errors (MSEs) are calculated through 1000 independent runs. The parameters of the algorithms were fine tuned to achieve their optimum performance at \SNRin = 0~dB and $f_{\mathrm{I}}=59$~Hz which is shown in (a). (b) In low  \SNRin, the proposed algorithm still yields low MSE, whereas the increased MSE using the other methods. (c) In high \SNRin, the proposed algorithm can further achieve lower MSE. (d) At $f_{\mathrm{I}}=56$~Hz; unlike the two other methods, the performance of the proposed algorithm is highly insensitive to the initial frequency deviations. As can be seen in (a)-(d) the proposed algorithm consistently yields faster convergence along with lower MSE compared with the other methods. The nominal frequency is set to 60~Hz in the Ziarani's and Martens' algorithms; however, the proposed algorithm does not require a priori setting of nominal frequency. Parameter setting, $\fs=1$~kHz, Proposed:\{$B_0=50$, $B_{\mathrm{st}}=0.5$, $B_{\infty}=0.1$, $P_0=0.1$, $P_{st}=0.5$, $P_{\infty}=1$, $W=1$\}, Ziarani: \{$\mu_1=8$, $\mu_2=1000$, $\mu_3=0.02$\}, Martens:\{$\tau=200$, $\zeta =1$, $\omega_n/\omega_p^n=0.02$\}}
\end{figure*}

\subsection{Comparison with Other Methods} \label{sec:comparison}

The algorithm is compared with narrow- and wide-band notch filtering, and two adaptive algorithms proposed by Ziarani et~al. \cite{ziarani_nonlinear_2002} and Martens et~al. \cite{martens_improved_2006}.

A performance comparison in terms of SNR improvement, mean square error (MSE) and convergence speed is presented in \sref{sec:perfcompare}. A comparison between the effects of different interference removal methods on synthetic neural oscillations is made in \sref{sec:effectosc}.

\subsubsection{Performance Comparison} \label{sec:perfcompare}

\mbox{\Fref{fig:notchfilt}} shows the effect of wide- and narrow-band notch filtering on a synthetically corrupted ECoG signal. The interference fundamental frequency was slightly deviated from 60~Hz which translated to even higher deviations in higher harmonics (\mbox{\fref{fig:interf}}). As can be seen in \mbox{\fref{fig:narrownotch}}, narrow-band notch filters fail to adequately remove the interference with changing frequency. On the other hand, wide-band notch filters distort the signal PSD (\mbox{\fref{fig:widenotch}}). The result of interference cancellation using the proposed method is displayed in \mbox{\fref{fig:proposed}}. It can be seen that, the interference is adequately removed while the signal frequency bands are highly preserved.

The adaptive methods of Ziarani et~al. \cite{ziarani_nonlinear_2002} and Martens et~al. \cite{martens_improved_2006} have been widely applied to electrocardiography (ECG) signals and shown effective in removing non-stationary power line interference. Here, we compare the convergence behaviour and the asymptotic performances of these methods against the proposed algorithm.

The first simulation is done to evaluate the asymptotic performances of the algorithms in terms of \SNRout versus \SNRin . For this purpose, randomly selected portions of an interference-free ECoG recording were used and each of which was superimposed with interference containing a single stationary sinusoid at $f_{\mathrm{I}}=59$~Hz with a random phase and a determined amplitude. The algorithms were allowed to fully converge to their steady states and the values of \SNRout were calculated for t\;$>$\;60~s. As can be seen in \fref{fig:SNRin_SNRout_errbars}, for all \SNRin values, the proposed algorithm achieves significantly higher \SNRout compared with other methods. Furthermore, the small minimum and maximum deviations---shown by the error bars---indicate the reliable convergence and consistent performance of the proposed algorithm. In this simulation, large lower error bars indicate that the algorithm under test may fail to converge.

The second simulation is carried out to evaluate the convergence behaviour of the adaptive algorithms in the mean sense. For this purpose, For this purpose, random signals with $1/f$ PSD (mimicking neural signal PSD) were generated, each of which was superimposed with a single sinusoid (mimicking the interference) whose frequency was slightly deviated from 60~Hz. Subsequently, the MSEs between the output of each algorithm and the actual random signal (without the interference) were calculated. The simulation was then repeated with different values of \SNRin and interference frequency ($f_{\mathrm{I}}$). In this evaluation, faster convergence and lower MSE values are desirable factors. \Fref{fig:MSE_conv} shows that the proposed algorithm consistently achieves faster convergence and lower MSE compared with the other methods. Furthermore, it can reasonably achieve its optimum performance regardless of the initial deviation of the interference frequency from its nominal value.

In the simulations, we observed that the two other adaptive methods were sensitive to the large amplitude artefacts---which are usually present in neural recording---such as electrode displacement and movement artefacts. In addition, since the performance of the algorithms depend on their parameter setting, we fine tuned the parameters of each algorithm to achieve its best performance---in terms of lower MSE and faster convergence---at \SNRin = 0~dB and \mbox{$f_{\mathrm{I}}=59$~Hz}. Furthermore, the adaptation blocking in the Martens' algorithm is not applicable to ECoG signals, thus their SAC~2 method was used.

\begin{figure*}[]
	\centering
\captionsetup[subfigure]{oneside,margin={5mm,0mm}}
	\subfloat[][]{\includegraphics{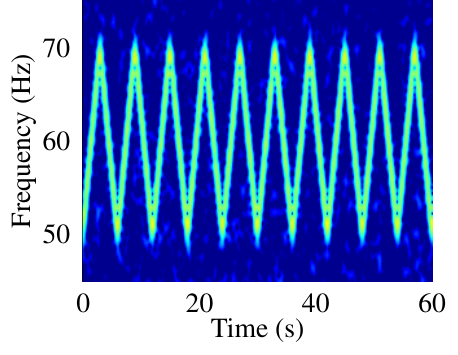}\label{fig:tf1}}
   	\subfloat[][]{\includegraphics{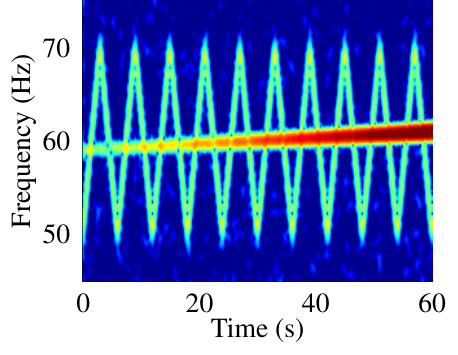}\label{fig:tf2}}
\captionsetup[subfigure]{oneside,margin={-8mm,0mm}}
	\subfloat[][]{\includegraphics{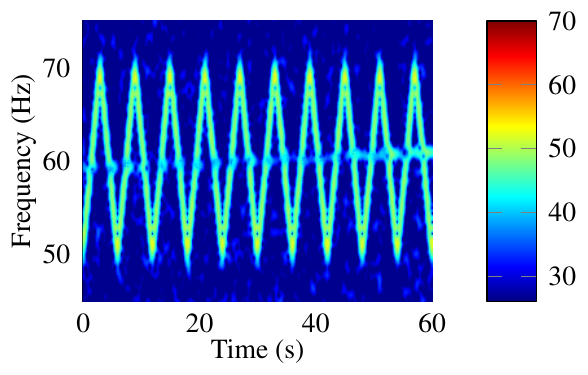}\label{fig:tf3}}
	\\
\captionsetup[subfigure]{oneside,margin={2mm,0mm}}
   	\subfloat[][]{\includegraphics{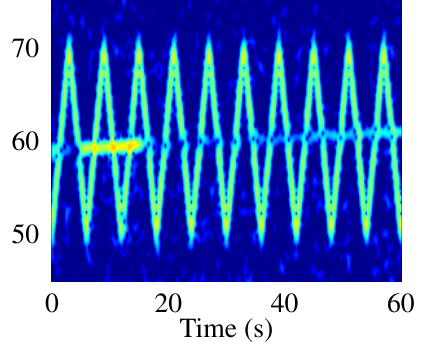}\label{fig:tf4}}
   	\subfloat[][]{\includegraphics{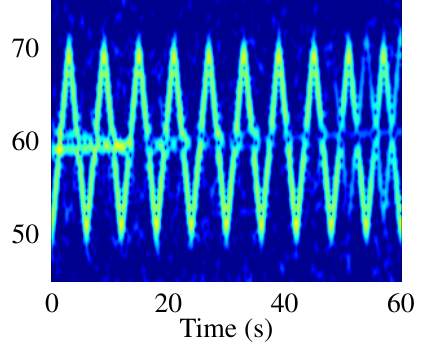}\label{fig:tf5}}
	\subfloat[][]{\includegraphics{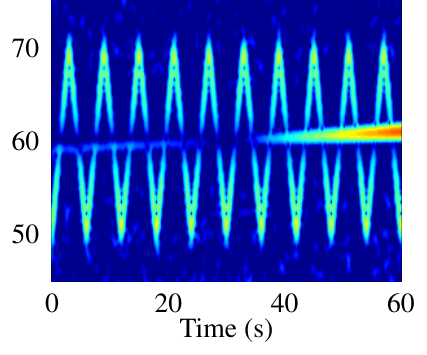}\label{fig:tf6}}
	\subfloat[][]{\includegraphics{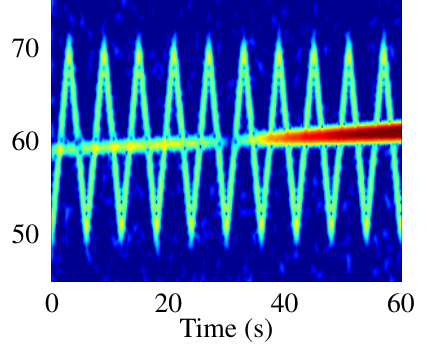}\label{fig:tf7}}
	\caption{Simulation with synthetic oscillations. Time-frequency plots: (a) Synthesized signal consisting of  bidirectional chirp between 50--70~Hz (representing the signal of interest) superimposed with a $1/f$ PSD background signal (representing neural noise). (b) After adding a sinusoidal interference whose frequency was swept from 59~Hz to 61~Hz, and its amplitude was logarithmically increased, setting \SNRin from 10~dB to $-$20~dB. (c) The proposed algorithm has tracked and removed the interference while reasonably preserving the signal components. (d) In the Martens' algorithm, the phase-locked loop (PLL) has become out of lock due to the oscillations (e.g.~5\;<\;t\;<\;15~s).
(e) The Ziarani's algorithm is sensitive to the interference power, thus same parameters cannot be used to obtain adequate performance for different power of the signal and/or interference. Furthermore, it has distorted the signal near the interference frequency band. (f) 10-Hz notch filter has excessively removed the signal components. (g) 1-Hz notch filter has only attenuated the interference near $t=30$~s when its frequency was close to 60~Hz, and failed to remove the interference otherwise.
The resultant \SNRout values (for $0<t<60$) are displayed in table~\ref{tab:tf_snr}. Parameter setting, $\fs=1$~kHz, Proposed:\{$B_0=20$, $B_{\mathrm{st}}=0.5$, $B_{\infty}=0.1$, $P_0=0.2$, $P_{st}=1$, $P_{\infty}=0.5$, $W=1$\}, Ziarani: \{$\mu_1=12$, $\mu_2=0.001$, $\mu_3=0.2$\}, Martens:\{$\tau=100$, $\zeta =1$, $\omega_n/\omega_p^n=0.01$\}\label{fig:time-freq}}
\end{figure*}

\subsubsection{Effects on Synthetic Oscillations} \label{sec:effectosc}
Neural oscillations (bio-markers) can appear at, or in the vicinity of, the interference frequency bands. Since these oscillations are useful for information decoding, it is important to ensure that they are well preserved and/or undergo minimal distortion during interference cancellation. To illustrate the performance of the algorithm in this regard, it is tested on synthetic oscillations contaminated with interference.
For this purpose, a sequence of patterned oscillations in the range of 50--70~Hz was generated and then added to a background random signal with \mbox{$1/f$}~PSD (\fref{fig:tf1}). This sequence represents a synthetic neural signal. Subsequently, a sinusoid (representing the interference) was synthesized and added to the signal. The frequency of this sinusoid was swept from 59~Hz to 61~Hz, and its amplitude was logarithmically increased, setting \SNRin from 10~dB to -20~dB (\fref{fig:tf2}).

Different methods of interference removal were applied to the synthesized signal. \Fref{fig:tf3} illustrates that the proposed algorithm has tracked and removed the interference while reasonably preserving the signal components. \Fref{fig:tf4} shows that in the Martens' algorithm, the phase-locked loop (PLL) has become out of lock due to the oscillations (e.g.~5\;<\;t\;<\;15~s).
\Fref{fig:tf5} shows that the Ziarani's algorithm is sensitive to the interference power, thus same parameters cannot be used to obtain adequate performance for different power of the signal and/or interference. Furthermore, it has distorted the signal near the interference frequency band. \Fref{fig:tf6} indicates that 10-Hz notch filter has excessively removed the signal components. \Fref{fig:tf7} shows that the 1-Hz notch filter has only attenuated the interference near $t=30$~s when its frequency was close to 60~Hz, and failed to remove the interference otherwise. 
The resultant \SNRout{s} (calculated for $0<t<60$) are displayed in table~\ref{tab:tf_snr}.

\begin{table}
\caption{\label{tab:tf_snr} Results of simulation with synthetic oscillations}
\centering
\begin{tabular}{lL{2,2}}
\toprule
Methods & \multicolumn{1}{c}{\SNRout (dB)} \\
\midrule
Proposed & 12.06 \\
Martens & 8.60 \\
Ziarani & 7.90 \\
10-Hz Notch & 2.20 \\
1-Hz Notch & -7.85 \\
\bottomrule
\end{tabular}
\end{table}

\subsection{Performance Evaluation on Real Data} \label{sec:realsim}
\begin{figure*}[t]
	\centering
	\captionsetup[subfigure]{oneside,margin={6mm,0mm}}
	\begin{flushleft}
	Extracellular:
	\vspace{-5mm}	
	\end{flushleft}
	     	\subfloat[][]{\label{fig:intpsdreal1}\includegraphics{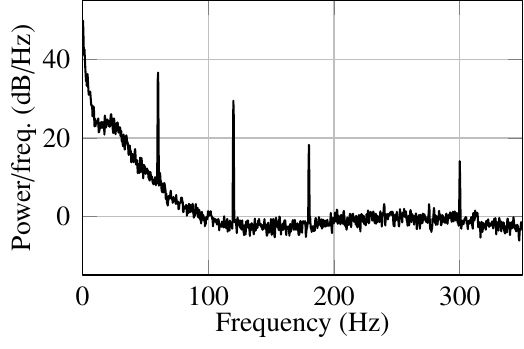}}
	    	\quad
		   \subfloat[][]{\label{fig:intpsdreal2}\includegraphics{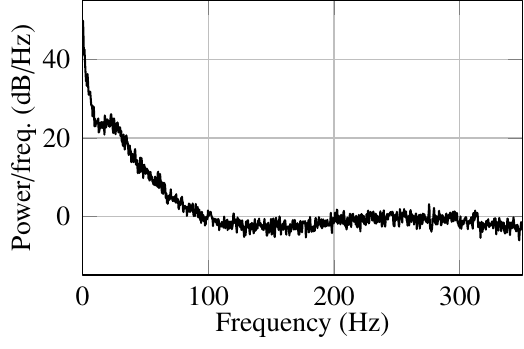}}
	    	\quad
	    	\subfloat[][]{\label{fig:intrealsig}\includegraphics{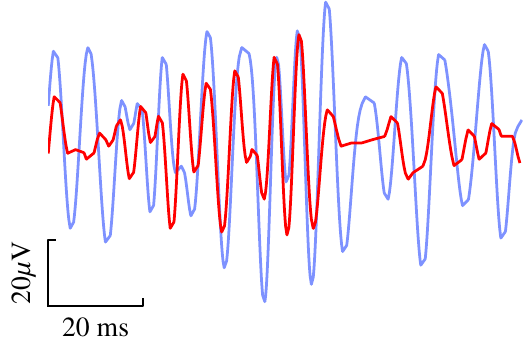}}
	\\
	\begin{flushleft}
	\vspace{-2mm}
	ECoG:
	\vspace{-5mm}		
	\end{flushleft}
	    	\subfloat[][]{\label{fig:ecogpsdreal1}\includegraphics{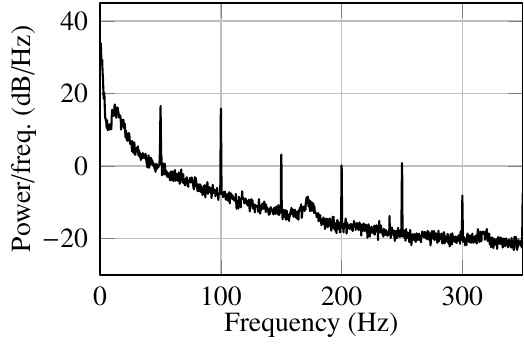}}
	    	\quad
		   \subfloat[][]{\label{fig:ecogpsdreal2}\includegraphics{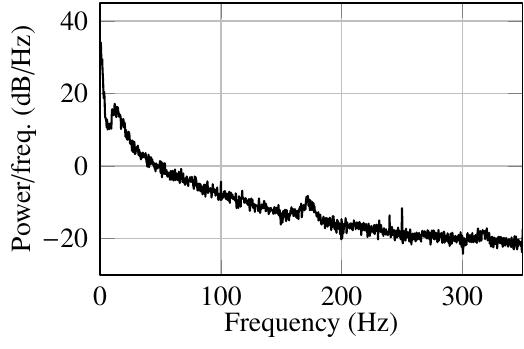}}
	    	\quad
	    	\subfloat[][]{\label{fig:ecogrealsig}\includegraphics{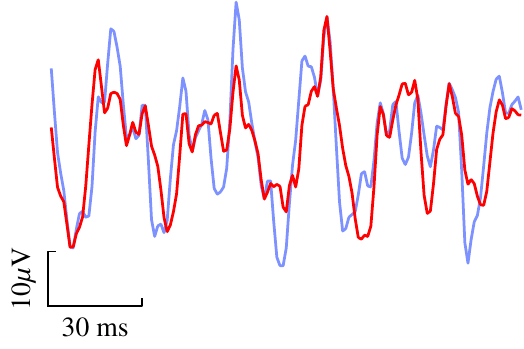}}
	\\
	\begin{flushleft}
	\vspace{-2mm}
	EEG:
	\vspace{-5mm}	
	\end{flushleft}
	    	\subfloat[][]{\label{fig:eegpsdreal1}\includegraphics{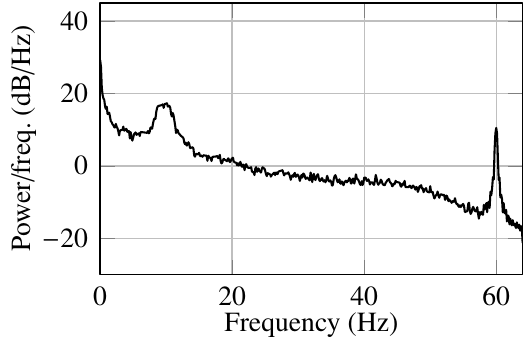}}
	    	\quad
		    \subfloat[][]{\label{fig:eegpsdreal2}\includegraphics{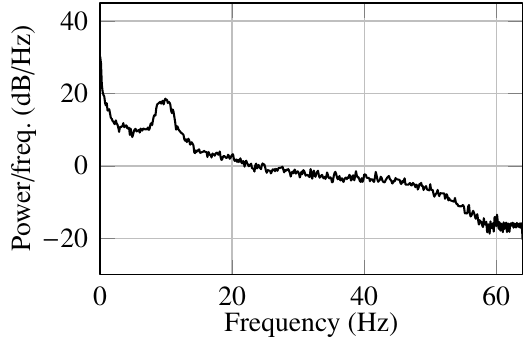}}
	    	\quad
	    	\subfloat[][]{\label{fig:eegrealsig}\includegraphics{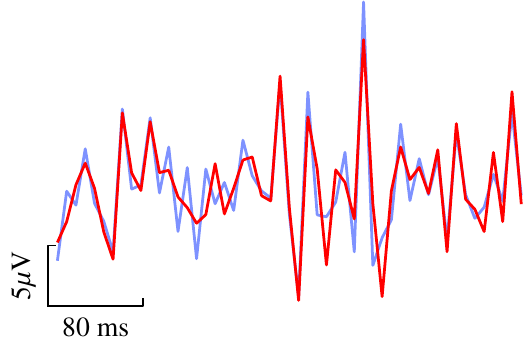}}
		\caption{Results of the experiment with real data. First column: PSD of the actual recorded signals containing power line interference. Second column: after applying the proposed algorithm, where the interference harmonics are removed. Third column: the actual recorded signal (\protect\plotref{plt:sigbefore}), and after interference cancellation by the proposed algorithm (\protect\plotref{plt:sigafter}), displayed in the gamma band ($>30$~Hz).
Note that in (b), the PSD remained minimally affected at 240~Hz (4\textsuperscript{th} harmonic), where no harmonic was present (compare with (a)).
In this experiment common parameter setting is  \{$B_0=50$, $B_{\infty}=0.05$, $B_{\mathrm{st}}=1$, $P_0=0.1$, $P_{\infty}=4$, $P_{st}=1$\}; and specific parameters settings are, extracellular: \{$\fs=40$~kHz, $W=2$\}, ECoG: \{$\fs=1$~kHz, $W=1$\}, EEG: \{$\fs=128$~Hz, $W=0.5$\}.}
	\label{fig:real}
\end{figure*}
Three types of biosignals including extracellular, ECoG and EEG recordings were used to demonstrate the performance of the algorithm on real data. These signals were recorded in ordinary environments, thus containing a significant amount of power line interference. The PSD of the recorded signals are displayed in figures~\ref{fig:intpsdreal1},~\subref*{fig:ecogpsdreal1}~and~\subref*{fig:eegpsdreal1} where the presence of the power line interference can be clearly seen. It can be observed that, the harmonics' power can be tens of dB higher than the signal power at the contaminated bands. Furthermore, both odd and even harmonics may be present in the recorded signals.

The extracellular, ECoG and EEG recordings were sampled at 40~kHz, 1~kHz and 128~Hz, respectively. The algorithm was applied to the recorded signals to cancel the interference. Figures~\ref{fig:intpsdreal2},~\subref*{fig:ecogpsdreal2}~and~\subref*{fig:eegpsdreal2} show the PSDs after interference cancellation, where the harmonics have been removed. In addition, the algorithm does not distort the signal frequency components where no harmonic is present. For example in \fref{fig:intpsdreal2}, the PSD remained unchanged at the frequency of the 4\textsuperscript{th} harmonic.
Figures~\ref{fig:intrealsig},~\subref*{fig:ecogrealsig}~and~\subref*{fig:eegrealsig} display portions of the gamma band signals before and after interference cancellation.

It is worth mentioning that, a favorable property of the proposed algorithm is that the proper parameter values remained the same (except $W$) for various signal modalities and sampling rates, confirming the usefulness of defining the alternative parameters. This property makes the algorithm easy to apply on various types of biopotential recordings, with only a slight adjustment of its parameters.

\section{Discussion} \label{sec:discussion}
In the design of the algorithm, a number of techniques are used to reduce its computational complexity. First, trigonometric function calculations are avoided in several parts including frequency estimation, harmonic frequencies calculation, and harmonic sinusoids generation. Second, the RLS algorithm is simplified by diagonal approximation of its covariance matrix.
These considerations are particularly important in hardware implementation, and significantly reduce the circuit area. To optimize the area further, a number of resource sharing techniques are used to share dividers, multipliers and some reusable circuit blocks at the cost of an increased clock frequency. In general, the circuit blocks of the discrete oscillator and the amplitude/phase estimator can be reused to remove $M$ number of harmonics at a $M$ times higher clock rate while requiring $7M'$ registers to store the state variables. Furthermore, the same circuitry can be reused to implement the two RLS update equations with 2 times higher clock rate. In this case, the blocks operate at different clock rates at $\fs$, $2\fs$, $M'\fs$ and $2M'\fs$.

The algorithm provides instantaneous interference cancellation, implying a zero phase shift between the input and output signals. In hardware implementation, the delay between input and output signals is determined by the circuit propagation delay which depends on several factors including the critical path of the circuit, fabrication technology and temperature. In our implementation in a 65-nm technology, this delay is negligible ($\ll 1\ \mu$s) considering the typical sampling rates used in biopotential recording (i.e.~up to tens of kHz).

Although the algorithm is mainly proposed for power line interference cancellation, it can also be used to cancel other types of harmonic interferences which may present in the recording. For this purpose, the corner frequencies of the bandpass filter should be adjusted accordingly. Furthermore, the algorithm is applicable to other types of biopotential recordings including ECG and electromyography (EMG) with no modification; however, the test results on these recordings are not presented in this paper. Moreover, we tested the algorithm on signals corrupted with different types of artefacts such as muscle, eye movement, electrode displacement, and other low and high frequency artefacts. We observed that the algorithm is highly robust to such artefacts. It is important to note that the input signal is assumed to be zero-mean, thus any DC bias should be removed before applying the algorithm.

It should be noted that the algorithm relies on the first harmonic of the interference for frequency estimation. In some situations, however, the first harmonic is suppressed by the recording amplifier, but higher harmonics are still present in the signal. In such situations, the bandpass filter can be accordingly adjusted ({e.g.}~to 90--130~Hz) to estimate the frequency of the second harmonic ($\kappa_2$), and the fundamental frequency estimate $\kf$ can be obtained through $\kf=\sqrt{(\kappa_2+1)/2}$ and subsequently be used for harmonic estimation.

\section{Conclusion} \label{sec:conclusion}
A robust and efficient algorithm is proposed to remove non-stationary 50/60~Hz interference and its harmonics, from neural recordings.
It is highly insensitive to the power of the interference, maintaining high output SNR (\textgreater~30~dB) in a wide range of signal and interference conditions. It can effectively track the variations in the frequencies, amplitudes and phases of the harmonics to cancel the interference without compromising the actual neural signals at the interference frequency bands.
Furthermore, it features low computational and memory requirements despite using no reference signal for estimation. This property makes the algorithm suitable for real-time applications and hardware implementation.

The convergence, tracking, and estimation accuracy of the algorithm can be controlled through several parameters. An alternative form of these parameters are introduced which have intuitive meaning, and make the parameter adjustment straightforward.

The performance of the algorithm is quantitatively evaluated in terms of output SNR, trade-off between settling time and \SNRout, and the convergence behaviour. High \SNRout (\textgreater~30~dB) is consistently achieved in different conditions of \SNRin ($-$30~dB to 30~dB), power line frequencies (45~Hz to 65~Hz), and sampling rates (as low as 100/120~Hz). Test results on the trade-off between settling time and \SNRout as well as the tracking behaviour,  demonstrate the fast adaptation of the algorithm to interference variations, while maintaining a high \SNRout. This makes the algorithm highly suitable for practical applications such as in wearable recording systems, where the interference power undergos large variations. Moreover, the algorithm features quick (\textless100~ms) initial convergence.

The comparative performance evaluation between the proposed algorithm with two other adaptive methods shows the improved performance of the proposed algorithm in terms of noise immunity, output SNR and convergence behaviour.

The algorithm is tested on real extracellular, ECoG and EEG recordings, where almost complete removal of the interference---while preserving the neural signals---is observed. It is also applicable to other types of biopotential recordings including ECG and electromyography (EMG), with no modification.

The MATLAB implementation of the proposed algorithm is provided at \cite{MATLABCode}, which caters to various biosignal recording applications. Furthermore, the chip implementation demonstrates the suitability of the algorithm for ASIC and real-time applications.

\ack
The ECoG and EEG data were downloaded from \nolinkurl{http://neurotycho.org/} \cite{shimoda_decoding_2012} and \nolinkurl{http://physionet.org/} \cite{goldberger_physiobank_2000}, respectively.
The authors would like to acknowledge Edward Keefer at Plexon, and Victor Pikov at Huntington Medical Research Institute for providing extracellular recording data.

The authors would also like to thank lab members Tong Wu for assisting in chip design and testing, and Azam Khalili and Amir Rastegarnia for their comments on an earlier draft of this paper.

The authors acknowledge the funding support by A*STAR PSF Grant R-263-000-699-305, A*STAR-CIMIT Grant R-263-000-A32-305, and NUS Grant R-263-000-A29-133.

\appendix
\section{Mathematical Derivations} \label{sec:apxMath}
\setcounter{section}{1}
\subsection{RLS algorithm} \label{sec:apxRLS}
The RLS algorithm is used to adapt the weights of the adaptive linear combiner in \fref{fig:combiner}. The weighted least squares cost function is defined as
\begin{eqnarray}
    \label{eq:RLSerr}
    \eqalign{
	e_k(i) = x(i) - \hat{h}_k(i), \\
    E_k = \sum_{i=0}^{n} { \lambdaa^{n-i} e_k^2(i)},
    }
\end{eqnarray}
where $e_k(i)$ is the instantaneous error and $0<\lambdaa \ll 1$ is the forgetting factor. The RLS algorithm is described as follows. Let $\mathbf U_k(n) = [u_k(n) \,\, u'_k(n)]^{\mathrm{T}}$ be the input sample vector and $\mathbf W_k(n) = [\hat{b}_k(n) \,\, \hat{c}_k(n)]^{\mathrm{T}}$ be the parameter vector. The input sample correlation matrix is defined as
\begin{subequations}
\begin{eqnarray}
	\eqalign{
	\mathbf R_k(n)  &= \sum_{i=0}^{n} {\lambdaa^{n-i} \mathbf U_k(n) \mathbf U_k}^T(n) \\
	&=
	\left[ \begin{array}{cc}
		r_{1,k}(n) & r_{2,k}(n)	\\
		r_{3,k}(n) & r_{4,k}(n)	
	\end{array} \right].
	}	\label{eq:RLScorr}
\end{eqnarray}
$\mathbf R_k(n)$ can be recursively calculated through
\begin{eqnarray}
    \label{eq:RLSupdate}
    \eqalign{
	\mathbf R_k(0) = \epsilon I_2, \epsilon>0 \quad \mathrm{for} \; k=1, \cdots, M',\\
	\mathbf R_k(n) = \lambdaa \mathbf R_k(n-1) + \mathbf U_k(n) \mathbf U_k^{\mathrm{T}}(n).
    }
\end{eqnarray}
Now, $\mathbf W_k$ can be adapted through the following RLS update equation \cite{Farhang-Boroujeny_adaptive_1999}.
\begin{eqnarray}
    \mathbf W_k(n+1) = \mathbf W_k(n) - \mathbf R_k^{-1} \mathbf U_k(n) e_k(n)
\end{eqnarray}
\end{subequations}
In the standard RLS method, matrix inversion lemma is used to obtain $\mathbf R^{-1}_k$ in order to avoid inverse matrix calculation. In this work, we suggest a simplification on $\mathbf R_k$ which leads to a less computational parameter adaptation.
It is assumed that the forgetting factor parameter $\lambdaa$ is selected close to the recommended values (i.e $0.5{\,<\,} W {\,<\,}5$). In this case, it is shown in~\ref{sec:apxRLSsimp} that $r_{2,k}$ and $r_{3,k}$ would become very small compared with $r_{1,k}$ and $r_{4,k}$, thus they can be neglected and $\mathbf R_k$ becomes a diagonal matrix. This leads to simplified update equations as
\begin{eqnarray}
	\eqalign{
    r_{1,k}(-1) = r_{1,k}(-1) = \hat{b}_k(-1) = \hat{c}_k(-1) =0, \\
	r_{1,k}(n) = \lambdaa r_{1,k}(n-1) + u_k(n)^2 ,	\\
	r_{4,k}(n) = \lambdaa r_{4,k}(n-1) +  u'_k(n)^2 ,	\\
	\hat{b}_k(n) = \hat{b}_k(n-1) + u_k(n)e_k(n) / r_{1,k}(n) , \\
	\hat{c}_k(n) = \hat{c}_k(n-1) + u'_k(n)e_k(n) / r_{4,k}(n) .	
    }
\end{eqnarray}
These update equations are much simpler than that of \eref{eq:RLSupdate} with regards to the number of arithmetic operations.
We also investigated the effect of this simplification on the performance of the algorithm by comparing the results with the case that the standard RLS algorithm was used, and no significant difference in performance was observed, which validates the proposed approximation.

\subsection{Simplification of the RLS algorithm} \label{sec:apxRLSsimp}
We propose a simplification on the RLS algorithm used in the phase and amplitude estimation stage of the algorithm. This simplification is based on approximating the RLS sample correlation matrix $\mathbf R_k(n)$ in \eref{eq:RLScorr} with a diagonal matrix. The elements of $\mathbf R_k(n)$ can be expanded as
\begin{subequations}
\begin{eqnarray}
r_{1,k}= \sum_{i=0}^{n} \lambdaa^{n-i} v^2 \sin^2(k\wf i),
\end{eqnarray}
\begin{eqnarray}
r_{4,k}= \sum_{i=0}^{n} \lambdaa^{n-i} v'^2 \cos^2(k\wf i),
\end{eqnarray}
\begin{eqnarray}
r_{2,k}= r_{3,k}= \sum_{i=0}^{n} \lambdaa^{n-i} v v' \sin(k\wf i) \cos(k\wf i).
\end{eqnarray}
\end{subequations}
If it can be shown that for typical parameters values and sampling rates the coefficients $|r_{2,k}|$ and $|r_{3,k}|$ are much less than $|r_{1,k}|$ and $|r_{4,k}|$, then the sample correlation matrix $R_k(n)$ can be well-approximated by a diagonal matrix (i.e.~$r_{2,k}$= $r_{3,k}$=0). This subsequently leads to much less computational RLS update equations. In the following derivations, we first show that $|r_{2,k}|, |r_{3,k}| \ll r_{1,k}$; the inequality $|r_{2,k}|, |r_{3,k}| \ll r_{4,k}$ can be similarly derived and is not presented here.
We would like to show that the following inequality holds for typical parameter values:

\begin{figure}[t]
	\centering
    {\scriptsize \includegraphics{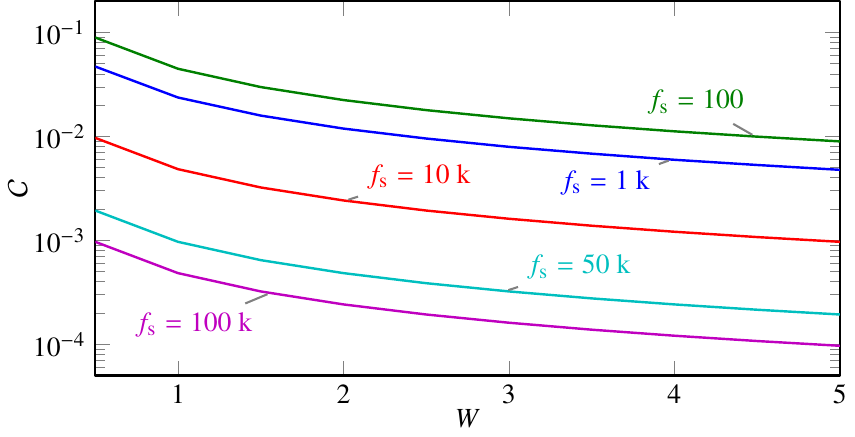}}
    \caption{The values of $\mathcal{C}$ at different sampling rates. It can be seen that at the minimum sampling rate $\fs=100$, the maximum value of $\mathcal{C}$ is less than $0.1\ll1$. The value of $\mathcal{C}$ monotonously decreases with increasing $\fs$ and $W$   }\label{fig:cvals}
\end{figure}

\begin{figure*}[!t]
    \renewcommand\thefigure{B1}
	\centering
   \includegraphics[width=\textwidth]{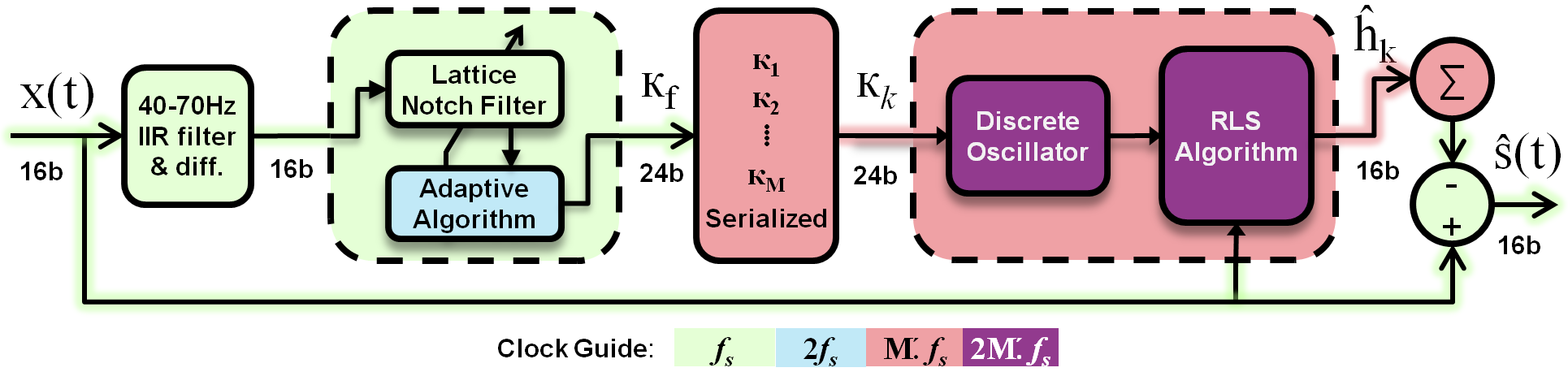}
    \caption{System architecture for implementation of power line interference cancellation algorithm. The system operates at multiple clock rates which are indicated in different colours. The slowest clock is equal to the signal sampling rate $\fs$ and the fastest clock depends on the number of harmonics to be removed and is equal to $2M'\fs$.}\label{fig:HWdiag}
\end{figure*}

\begin{subequations}
\begin{eqnarray}
	\eqalign{
	\left| \sum_{i=0}^{n} \lambdaa^{n-i} v v' \sin(k \wf i) \cos(k \wf i) \right|
	\ll \sum_{i=0}^{n} \lambdaa^{n-i} v^2 \sin^2(k\wf i) ,
	} \nonumber \\ \label{eq:inequ1}
\end{eqnarray}
or equivalently
\begin{eqnarray}
\frac{v v'}{v^2} \frac{\left| \sum_{i=0}^{n} \lambdaa^{n-i} \sin(k \wf i) \cos(k \wf i) \right|}{ \sum_{i=0}^{n} \lambdaa^{n-i} v^2 \sin^2(k\wf i)} \ll 1 . \label{eq:inequ1frac}
\end{eqnarray}
\end{subequations}
Note that the right-hand side of \eref{eq:inequ1} is always positive and equal to its absolute value. The magnitudes $v$ and $v'$ are not initially included for the following derivation and will be considered in the last stage. After a little manipulation of \eref{eq:inequ1} using trigonometric identities we get
\begin{eqnarray}
\frac{\sqrt{2}}{2} \left| \sum_{i=0}^{n} \lambdaa^{n-i} \sin(2k\wf i{+} \frac{\pi}{4}) \right| \ll \frac{1}{2}\sum_{i=0}^{n} \lambdaa^{n-i}
\end{eqnarray}
which can be expressed in the complex domain as
\begin{eqnarray}
	\eqalign{
	\sqrt{2} \cdot \left| \mathcal{I}m \left\{ e^{{\mathbf j}\frac{\pi}{4}}\frac{\lambdaa^{n+1}-e^{\mathbf{j} 2k\wf(n+1)}}{\lambdaa-e^{\mathbf{j} 2 k \wf}} \right\} \right|
	 \ll \frac{\lambdaa^{n+1}-1}{\lambdaa-1} .
	 }
\end{eqnarray}
Using the property $|\mathcal{I}m\{z\}| \le |z|$, where $z$ is a complex number, we can alternatively show that
\begin{eqnarray}
\eqalign{
\sqrt{2} \cdot \left| e^{\mathbf{j}\frac{\pi}{4}} \frac{\lambdaa^{n+1}-e^{\mathbf{j} 2k\wf(n+1)}}{\lambdaa-e^{\mathbf{j} 2 k \wf}} \right|= \\
\left[{2 \frac{1{+}\lambdaa^{2(n{+}1)}{-}2\lambdaa^{n{+}1}\cos(2 k \wf (n{+}1))}{1+\lambdaa^2 - 2\lambdaa\cos(2 k \wf)}}\right]^\frac{1}{2}  \ll \frac{\lambdaa^{n+1}- 1}{\lambdaa-1} .
}
\end{eqnarray}
We define
\begin{eqnarray}
	\mathcal{A} =
	\frac{\displaystyle{\left[{2 \frac{1{+}\lambdaa^{2(n{+}1)}{-}2\lambdaa^{n{+}1}\cos(2 k \wf (n{+}1))}{1{+}\lambdaa^2{-}2\lambdaa\cos(2 \wf)}}\right]^\frac{1}{2}}}
	{\displaystyle{\frac{\lambdaa^{n+1}-1}{\lambdaa-1}}}, \\
	\mathcal{B} = \max\{\frac{v}{v'},\frac{v'}{v}\}, \\
	\mathcal{C} = \max_{k,n}{\mathcal{A} \cdot \mathcal{B}},
\end{eqnarray}
where $v/v'$ is given by (see \cite{turner_recursive_2003})
\begin{eqnarray}
	\frac{v}{v'}= \sqrt{\frac{1+\cos(k\wf)}{1-\cos(k\wf)}}.
\end{eqnarray}

If we can show that $\mathcal{C} \ll 1$, then \eref{eq:inequ1frac} holds. To show this, numerical simulation is used to obtain the upper bound values on all the harmonics and in the operating condition where $n\gg1$. For this purpose, the values of $\lambdaa$ are obtained from $W$ through \eref{equ:lamset}, where $W$ is swept in the range of \mbox{0.5--5} which covers the recommended range. \Fref{fig:cvals} displays the values of $\mathcal{C}$ in different sampling rates. As can be seen, in all the conditions, $\mathcal{C}<0.1\ll1$ indicating that the inequalities \mbox{$|r_{2,k}|, |r_{3,k}| \ll r_{1,k}$} and $|r_{2,k}|, |r_{3,k}| \ll r_{4,k}$ hold.

\subsection{Forgetting factors, Settling time, and Notch bandwidth}
Here, we describe the relation between forgetting factors and settling time as well as the relation between pole radii and notch bandwidth. The proper values of the forgetting factors depend on the sampling rate, making it difficult to adjust their values in general condition. On the other hand, settling time is independent of the sampling rate and has a more intuitive meaning that makes the parameter tuning straightforward. In certain parts of the algorithm such as frequency estimation and phase/amplitude adaptation, settling time can be associated with the forgetting factor by the following formulation
\begin{eqnarray}
	0.95\frac{1}{1-\lambda}=\frac{1-\lambda^{n_\mathrm{set}+1}}{1-\lambda} \\
	\Rightarrow \lambda = \exp{{\frac{\ln(0.05)}{t_\mathrm{set}\fs+1}}}, \label{equ:APlam-settime}
\end{eqnarray}
where $n_\mathrm{set}=\fs t_\mathrm{set}$,  $\lambda$ is the forgetting factor, $t_\mathrm{set}$ is the desired settling time, and $\fs$ is the sampling rate. The transformation of \eref{equ:APlam-settime} is used in \eref{equ:lamset} to adjust $\alpha_{\mathrm{st}}$,  $\lambda_0$, $\lambda_{\mathrm{st}}$, $\lambda_{\infty}$, and $\lambdaa$.

Notch bandwidth is independent of the sampling rate and can be alternatively adjusted instead of the pole radii. Given the notch bandwidth $B$, the pole radii $\alpha$ is obtained by
\begin{eqnarray}
	\alpha = \frac{1-\tan{\pi B/\fs}}{1+\tan{\pi B/\fs}}. \label{equ:APnotchBW}
\end{eqnarray}

\begin{figure}[!t]
    \renewcommand\thefigure{B2}
	\centering
	\includegraphics[width=6cm]{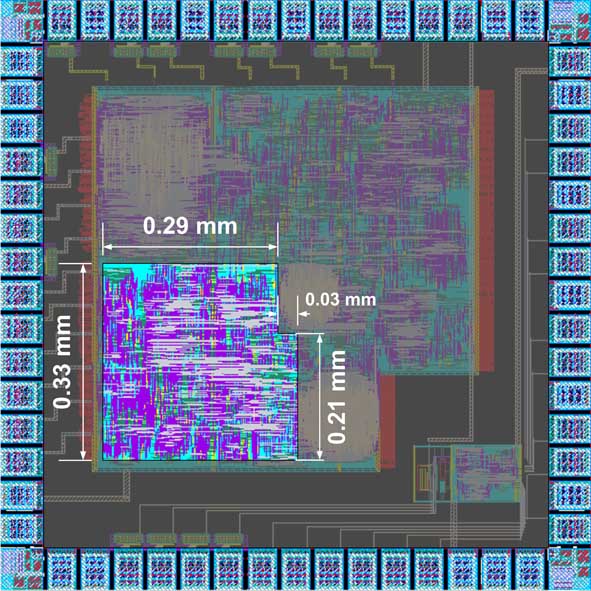}
    \vspace{2pt}
    \caption{The chip layout photo. The area consumed by the interference removal module is approximately 0.11~mm$^2$.}\label{fig:chip_layout}
\end{figure}

\section{Hardware Implementation} \label{sec:apxHardware}
\setcounter{section}{2}
\setcounter{figure}{2}
\begin{figure*}[!t]
	\centering
    \captionsetup[subfigure]{oneside,margin={6mm,0mm}}
    \subfloat[][]{\includegraphics{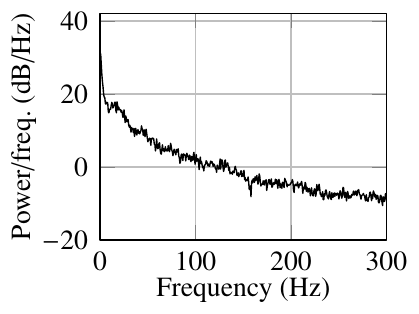}}\;
    \subfloat[][]{\includegraphics{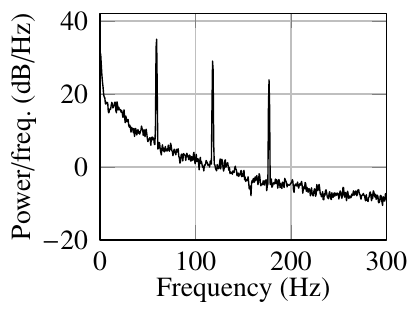}}\;
    \subfloat[][]{\includegraphics{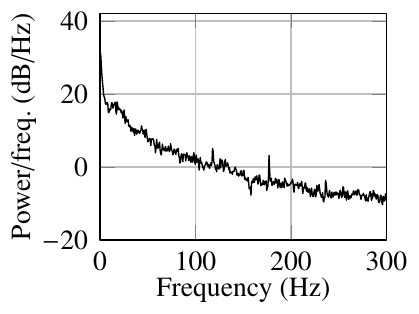}\label{fig:HW_PSDc}}\;
    \subfloat[][]{\includegraphics{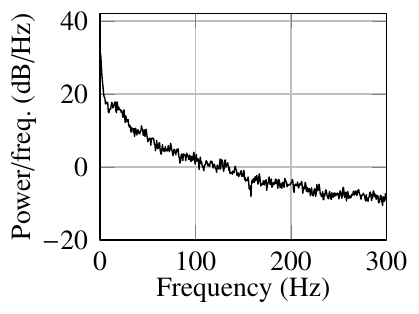}}\;
    \\
    \vspace{-15pt}
    \captionsetup[subfloat]{captionskip=-10pt}
    \subfloat[][]{\includegraphics{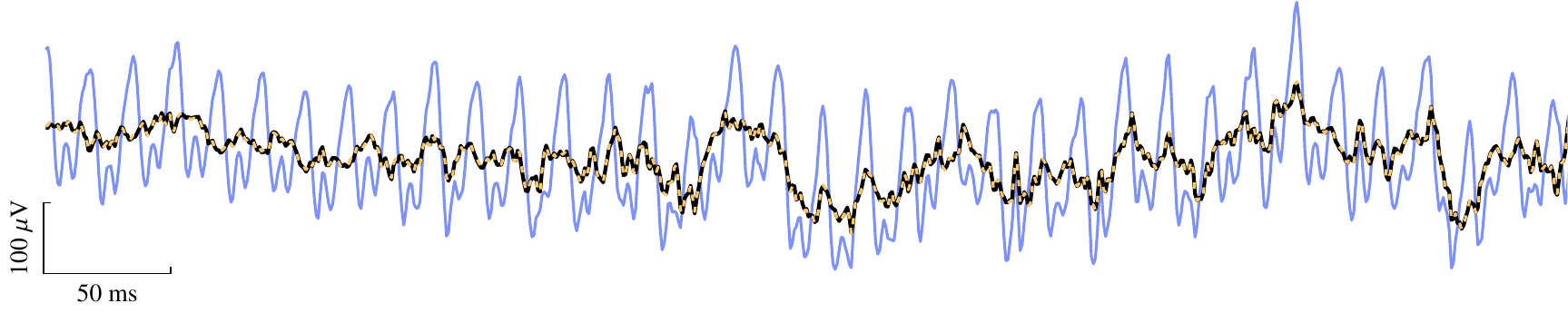}}
    \vspace{-2pt}
    \caption{Testing result with real-time data. (a) PSD of the clean ECoG signal used for simulation. (b) PSD of the signal contaminated with synthetic interference with the fundamental frequency of 59~Hz, which served as the chip input signal (\SNRin = 0~dB). (c) PSD of the chip output signal (\SNRout = 28.9~dB). (d) PSD of the output from the full-precision reference model (\SNRout = 31.1~dB). The slight difference between (c) and (d) is due the precision loss in fixed-point implementation. (e)   Plots of contaminated signal (\protect\plotref{hwplot1}), chip output (\protect\plotref{hwplot2}) and reference model output (\protect\plotref{hwplot3}), where the interference is removed, and the chip output accurately follows the reference model output.}\label{fig:HW_PSD}
\end{figure*}

\begin{figure*}[!t]
	\centering
    \includegraphics{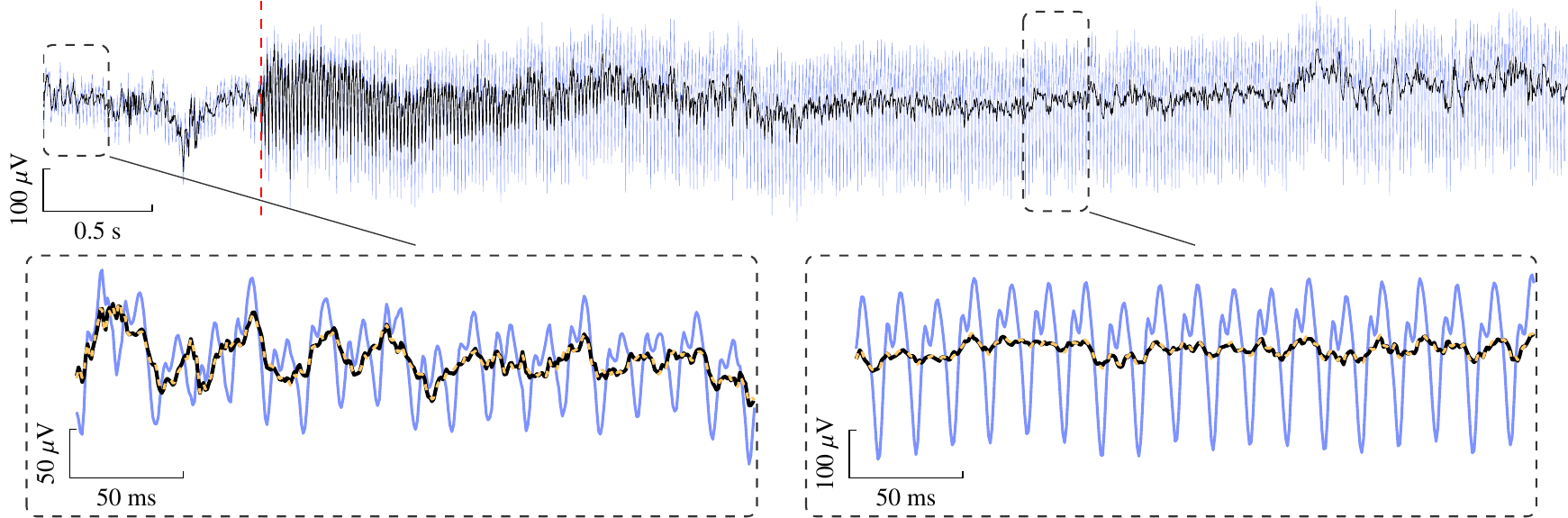}
    \caption{Response to a step change in interference power. The chip input signal (\protect\plotref{chipin}) was synthesized by adding a synthetic interference, containing three harmonics, to a pre-recorded clean ECoG signal. The power of the increased at the time instance indicated by (\ref{vertline}) (\SNRin = 0~dB before (\protect\plotref{vertline}) and \SNRin = $-$10~dB after (\protect\plotref{vertline})). The plots show the contaminated input signal (\protect\plotref{chipin}), the chip output signal (\protect\plotref{chipout}) and the full-precision reference model output (\protect\plotref{GMout}), where the interference is removed, and the chip output closely follows the reference model output.}\label{fig:HW_amp_sig}
\end{figure*}

\begin{figure*}[!t]
	\centering
	\includegraphics{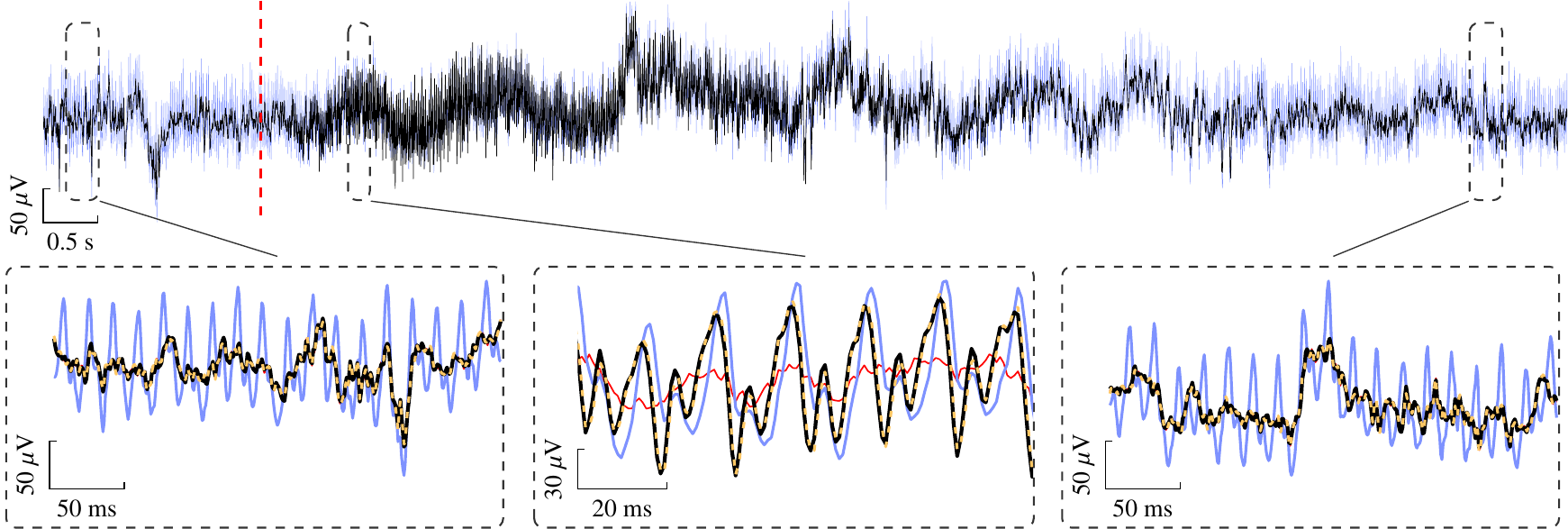}
    \caption{Response to a step change in the interference fundamental frequency. The fundamental frequency is changed from 60~Hz to 60.2~Hz at the time instance indicated by (\ref{vertline}). The plots show the original signal (\ref{orgsig}), the contaminated signal serving as the chip input (\protect\plotref{chipin}), the chip output signal (\protect\plotref{chipout}) full-precision reference model output (\protect\plotref{GMout}). It can be seen that the chip output adapts to the change and  closely follows the output of the reference model. }\label{fig:HW_freq_sig}
\end{figure*}

A prototype of the proposed algorithm was fabricated in a 65-nm CMOS process. All the circuit modules are coded in Verilog and implemented in fixed-point arithmetic. The implementation achieves real-time performance with instantaneous interference cancellation (only negligible propagation delay). To optimize the area, a number of resource sharing techniques are used to share dividers, multipliers and some reusable circuit blocks, at the cost of an increased clock frequency by at least 2 times the sampling rate. To account for higher harmonics, same circuit blocks are reused through multiplexing the inputs; hence, considerably saving the circuit area. In general, the cost for processing $M'$ harmonics is $7M'$ number of 24-bit registers as well as an increased system maximum clock rate of $2M'\fs$. The system requires four clock inputs with frequencies $\fs$, $2\fs$, $M'\fs$ and $2M'\fs$.

In the hardware implementation we set $\fs=1250$~Hz and $M'=3$ (i.e.~removing the first three harmonics), hence requiring a maximum clock rate of $7.5$~kHz. The input/output signals are represented by 16-bit signed integers. The clock rates and word lengths of the hardware modules are shown in \fref{fig:HWdiag}. The chip layout photo is displayed in \fref{fig:chip_layout}.

The functionality of the chip was verified against a reference model which was implemented in MATLAB and used full-precision arithmetic. The chip input signal was synthesized by adding a pre-recorded interference-free ECoG signal with a synthetic interference. We carried out three tests to demonstrate the chip functionality. In all the tests, the lengths of the input signals were 1~minute, and the \SNRout values were calculated for $t>20$~s.

In Test~1, \SNRin = 0~dB and the interference consisted of three stationary harmonics with the fundamental frequency of 59~Hz. The signal was then fed into the chip in real-time and the output was recorded. \Fref{fig:HW_PSD} shows the result of Test~1. The PSD of the chip output signal can be seen in \fref{fig:HW_PSDc} showing that the interference is significantly attenuated. Furthermore, the signals traces in the time domain clearly show that the chip output properly follows the reference model output.

In Test~2, the interference power was suddenly increased, changing the \SNRin from 0~dB to $-10$~dB. \Fref{fig:HW_amp_sig} shows the chip input and output signals. It can be seen that, the chip output consistently follows the reference model output. Moreover, after the step jump in the interference power at (\plotref{vertline}), the chip output adapted to the change to reject the residual amount of the interference.

In Test~3, \SNRin = 0~dB and the interference fundamental frequency was changed from 60~Hz to 60.2~Hz. The chip input and output signals are shown in \fref{fig:HW_freq_sig}. It can be seen that, the chip output consistently follows the reference model output. Moreover, after the frequency change at (\plotref{vertline}), the chip output adapted to the frequency change and approached the actual signal.

The \SNRout values are displayed in table~\ref{tab:chipSNR}. Slightly less \SNRout values of the chip output compared with the reference model output are due to the precision loss caused by fixed-point arithmetic used in the chip implementation.
\begin{table}[h]
\caption{\label{tab:chipSNR} Reference model and Chip \SNRout }
\centering
\begin{tabular}{lL{2,0}L{2,1}L{2,1}L{2,1}@{}}
\toprule
\multirow{2}{*}{Test} & \multicolumn{1}{c}{\multirow{2}{*}{\SNRin (dB)}} & \multicolumn{2}{c}{\SNRout (dB)}\\
\cmidrule(lr){3-4} & & \multicolumn{1}{c}{Reference model} & \multicolumn{1}{c}{Chip} \\
\midrule
Test 1 & 0 & 31.1 & 28.9 \\
Test 2 & -10 & 29.5 & 27.8 \\
Test 3 & 0 & 29.2 & 28.3 \\
\bottomrule
\end{tabular}
\resizebox {1pt}{1pt}{
\begin{tikzpicture}
    \begin{axis}[xmin=1,xmax=1,hide axis]
        \addplot [
        color=hwmycolor1,
        solid,
        line width=0.9pt
        ]
        (0,0);\label{hwplot1}
        \addplot [
        color=black,
        solid,
        line width=1.2pt
        ]
        (0,0);\label{hwplot2}
        \addplot [
        color=hwmycolor2,
        dash pattern=on 2pt off 3pt,
        line width=.9pt
        ]
        (0,0);\label{hwplot3}
        \addplot [
        color=red,
        solid,
        line width=0.5pt
        ]
        (0,0);\label{orgsig};
        \addplot [
        color=hwmycolor1,
        solid,
        line width=0.9pt
        ]
        (0,0);\label{chipin};
        \addplot [
        color=black,
        solid,
        line width=1.2pt
        ]
        (0,0);\label{chipout};
        \addplot [
        color=hwmycolor2,
        dash pattern=on 2pt off 3pt,
        line width=0.9pt
        ]
        (0,0);\label{GMout};
        \addplot [
        color=hwmycolor1,
        solid,
        line width=0.8pt
        ]
        (0,0); \label{plt:sigbefore}
        \addplot [
        color=red,
        solid,
        line width=0.8pt
        ]
        (0,0);  \label{plt:sigafter}
        \addplot [red,thick,dashed,forget plot] (0,0); \label{vertline}
        \addplot [
        color=black,
        dashed,
        line width=0.8pt
        ]
        (0,0); \label{plt:ampltrackdashed}
        \addplot [
        color=mycolor1,
        solid,
        line width = 0.5
        ]
        (0,0); \label{plt:orgint}
        \addplot [
        color=blue!80!black,
        solid,
        line width = 0.5
        ]
        (0,0); \label{plt:estint}
    \end{axis}
\end{tikzpicture}
}%
\end{table}

\section*{References}
\bibliographystyle{iopart-num}
\bibliography{allRefs,myurls}

\end{document}